\documentclass[aps, pre, twocolumn, showpacks, superscriptaddress, floatfix]{revtex4-1}

\usepackage{graphics}
\usepackage{epsfig}
\usepackage{array}
\usepackage{verbatim}
\usepackage{mathrsfs}
\usepackage{amsmath}
\usepackage{mathtools}

\newcommand{\be}{\begin{equation}}
\newcommand{\ee}{\end{equation}}
\newcommand{\bea}{\begin{eqnarray}}
\newcommand{\eea}{\end{eqnarray}}

\begin{document}

\title{Random Sequential Renormalization and Agglomerative Percolation in Networks:\\ Application to Erd\"os-R\'enyi and Scale-free Graphs}
\author{Golnoosh Bizhani} \affiliation{Complexity Science Group, University of Calgary, Calgary T2N 1N4, Canada}
\author{Peter Grassberger} \affiliation{Complexity Science Group, University of Calgary, Calgary T2N 1N4, Canada}
\author{Maya Paczuski} \affiliation{Complexity Science Group, University of Calgary, Calgary T2N 1N4, Canada}

\date{\today}

\begin{abstract}
We study the statistical behavior under random sequential renormalization (RSR) of several network models including Erd\"os-R\'enyi (ER) graphs, scale-free networks, and an annealed 
model (AM) related to ER graphs. In RSR the network is locally coarse 
grained by choosing at each renormalization step a node at random and joining it to all its neighbors.
Compared to previous (quasi-)parallel renormalization methods 
[Song {\it et. al.}, Nature (London) {\bf 433}, 392 (2005)], RSR allows a more fine-grained analysis 
of the renormalization group (RG) flow and unravels new features
that were not discussed in the previous analyses. In particular, we
find that all networks exhibit a second-order transition in their RG flow. 
This phase transition is associated with the emergence of a giant hub and 
can be viewed as a new variant of percolation, called agglomerative percolation. We claim that this transition exists
also in previous graph renormalization schemes and explains some of the scaling behavior seen 
there. For critical trees it happens as $N/N_0 \to 0$ in the limit of large systems (where $N_0$ is the 
initial size of the graph and $N$ its size at a given RSR step). In contrast, it happens at finite $N/N_0$
in sparse ER graphs and in the annealed model, while it happens for $N/N_0\to 1$ on
scale-free networks. Critical exponents seem to depend on the type of the graph but not on the 
average degree and obey usual scaling relations for percolation phenomena. For the annealed model 
they agree with the exponents obtained from a mean-field theory. 
At late times, the networks exhibit a star-like structure in agreement with the results of 
Radicchi et. al. [F. Radicchi {\it et. al.} Phys. Rev. Lett. {\bf 101}, 148701 (2008)].
While degree distributions are of main interest when regarding the scheme as network renormalization,
mass distributions (which are more relevant when considering ``supernodes'' as clusters) are much 
easier to study using the fast Newman-Ziff algorithm for percolation, allowing us to obtain very high
statistics.

\end{abstract}
\maketitle
\section{Introduction}

Complex networks provide a useful representation for complex phenomena in 
a variety of settings including social, biological and technological systems, and have 
been studied extensively in the past decade~\cite{pastor, newmanRev, boccaletti}. 
A common property of many complex real world networks is the heterogeneity of nodes leading
to wide (power law, ``scale-free") degree distributions~\cite{BA}. 

For systems embedded in Euclidean space, scale-free statistics is often related to the notion of 
self-similarity. In statistical physics and critical phenomena this is usually studied by 
using the renormalization group (RG) technique, where degrees of freedom of the system are 
eliminated successively by coarse-graining. The scaling behavior of the systems close to the 
fixed point of the RG flow is then examined and systems with similar scaling behavior are 
classified into universality classes~\cite{Stanley, Stauffer}. 

While renormalization is well defined and extensively studied for spatially extended systems
(including regular lattices and disordered systems), it is not clear whether it can be applied
to complex networks that have no spatial structure, where the topology is given only
by the network itself. Naively one would expect that the ``small world" property displayed
by many real networks~\cite{Milgram,Watts} means that they cannot be embedded in 
any finite dimensional space, and thus renormalization schemes should be less useful.
Nevertheless, a real space renormalization transformation for such networks was introduced by 
Song {\it et al.}~\cite{Song, Song3}. In this scheme, the entire network is covered in 
each RG step by a set of boxes, and each box is considered as a ``supernode" in the next 
RG step. Several complex networks were claimed to have a finite self-similar 
or fractal dimension, i.e. the number of boxes needed to cover the network seemed to show a 
power-law relation with the diameter of the box, in blatant contradiction to their small 
world property. Although this issue was never solved, it was suggested that the fractality 
of real world networks depends on self organization in the growth mechanism~\cite{Song2}, 
assortativity of fractal networks~\cite{Yook05}, and fractality of their underlying 
structure~\cite{Kim1, Kim2, Kim3, Goh06, Song10}. 

This conflict between the ``small-world" property and any fractality of complex networks 
was avoided by Radicchi {\it et al.}~\cite{Radi1, Radi2} by using an RG analysis based on 
the same box covering idea, but studying carefully the RG flow itself, without using any 
length scale dependence for making claims about fractality.

There are some technical concerns in these previous box covering methods for 
renormalizing networks. First, according to the original idea of Hausdorff~\cite{Falconer}, 
the sizes of boxes should be individually optimized, whereas in the suggested methods all 
boxes are of equal size. This is a particularly severe problem due to the heterogeneous 
connectivity in complex networks that leads also to very wide distributions of nodes per box,
most of them being nearly empty.
Secondly, even when boxes of the same size are used, the precise placement of boxes strongly affects 
 the result, and optimizing their positions is not practically feasible. Although 
the suggested methods in Refs.~\cite{Song, Song2, Song3, Kim1} are claimed to overcome this 
problem, their results still depend on the order at which the boxes are laid down, making
these schemes quasi-sequential. In particular, the number of nodes per box decreases strongly with
the number of boxes already put down. Finally, during each RG step the size of the network decreases 
dramatically, which results in a small number of data points in the RG flow. For networks with 
small-world property this is particularly serious, as the diameter of the networks scales only 
with $\log(N)$ ($N$ being the size of the network). To compensate this, only parts of the network 
have been coarse-grained in Ref.~\cite{Song10} at each step of renormalization, which 
adds more complexity to the process and makes the results even more difficult to interpret.

In our previous work~\cite{tree_paper}, we suggested a completely sequential renormalization 
scheme for undirected and unweighted graphs, called Random Sequential Renormalization (RSR). 
In RSR at each step of renormalization one node is chosen randomly, and all nodes within a given 
distance $b$ are replaced by a single super-node. All links from the outside to the (removed) 
neighborhood are redirected to the super-node, and the super-node is then treated like any other 
node in the network. The parameter $b$ is called the box radius. 

RSR has the advantage that it does not involve any optimum tiling and is very easy to code and 
understand. It avoids the problem of mostly-empty boxes. Furthermore, as the network is affected only 
locally and the decimation is considerably less at each step of RSR, the whole flow generates 
much more statistics which allows a more detailed analysis. 

Another advantage of RSR is that it can be interpreted as a cluster growth process, where 
initially all nodes are considered to be clusters of mass one. At each step of RSR a randomly 
chosen cluster grows by agglomerating with all its neighboring clusters.  Using the fast Monte Carlo 
algorithm for percolation introduced by Newman and Ziff (NZ)~\cite{newman_ziff1, newman_ziff2}, RSR 
can be easily implemented on networks with millions of nodes.

In our first paper on RSR~\cite{tree_paper}, we applied this method to critical trees. 
Their simple structure makes it possible to study the renormalization flow analytically, giving perfect agreement with results from numerical simulations. 
We found three regimes in the evolution of critical trees under RSR: (i) An initial regime with 
small fluctuations in the region $N_0^{1/2}\lesssim N<N_0$ (with $N_0$ being the initial size 
of the network and $N$ its size at a given renormalization step); (ii) An intermediate regime for 
$N_0^{1/4}\lesssim N \lesssim N_0^{1/2}$ where the network is a fat, short tree whose structure 
is dominated by a giant hub. The transition between these two regimes is associated with 
emergence of a power-law degree distribution and is described by crossover functions exhibiting 
finite-size scaling; (iii) A third regime extending down to $N=1$ where the network is 
a star with a central hub and many leaves.

The appearance of power-law distributions and scaling is associated with a continuous transition, 
called ``agglomerative percolation" (AP)~\cite{claireAP}. In one dimension (i.e. graphs 
consisting of a simple 1-d chain), AP has been solved exactly~\cite{sw1D, swmassdep}. There it shows 
non-trivial scaling with exponents that depend on the box size~\cite{sw1D}. In two dimensions,
AP is for triangular lattices in the same universality class as ordinary percolation (OP), whereas 
it shows different critical behavior for square lattices~\cite{claireAP}. This is related to the bipartite structure of the square lattice as every site on the boundary of any cluster is on the same sub lattice~\cite{H-W}. The fact that patently 
non-fractal structures like one and two dimensional lattices also exhibit scaling under RSR suggests that some of the scaling laws 
previously found in small-world networks are due to agglomerative percolation transition, rather than any underlying fractality of most networks.

In the present paper we study the behavior of sparse Erd\"os R\'enyi (ER) graphs and of the 
scale-free model of Barab\'asi and Albert~(BA)~\cite{BA} under RSR. For sparse ER graphs under 
RSR with $b=1$, we find a continuous percolation transition at finite $x=N/N_0$. Using finite 
size scaling methods, we show that the corresponding critical exponents are consistent with a 
scaling theory based on two independent exponents. Within our error estimates, these exponents 
appear to be independent of the initial average degree of the ER graphs. For the BA model the transition seems to be pushed to $x=1$, which makes it more difficult to obtain precise numerical results.

We also study RSR analytically using a mean-field theory based on generating functions. 
The behavior of graphs before the AP transition is consistent with this theory. After the 
transition the theory fails due to large fluctuations, as well as due to the effect of loops that 
are negligible before the transition. The predictions of the 
theory are in agreement with our simulations of an annealed model.

We introduce our model and simulation method in Sec.~II, where we also define the graphs and the sizes of the ensembles under study.  
Section~III presents our simulation results for ER graphs. We show evidence of a continuous 
percolation transition, find the scaling properties and the corresponding critical exponents 
numerically, and show that they obey the common scaling relations of ordinary percolation. In Sec.~IV we 
develop a mean-field theory for the evolution of ER graphs under RSR, and compare its results 
with our simulations of an annealed model.
We discuss the behavior of graphs beyond the percolation transition in Sec~V. 
Sections~VI and VII are devoted to the results of simulations on ER graphs with different 
average degrees as well as RSR with larger box sizes. Finally in Sec.~VIII we examine BA networks, and we conclude our study in Sec.~IX.

\section{The model}
\subsection{Random Sequential Renormalization}
\begin{figure}
\includegraphics[width=1\columnwidth]{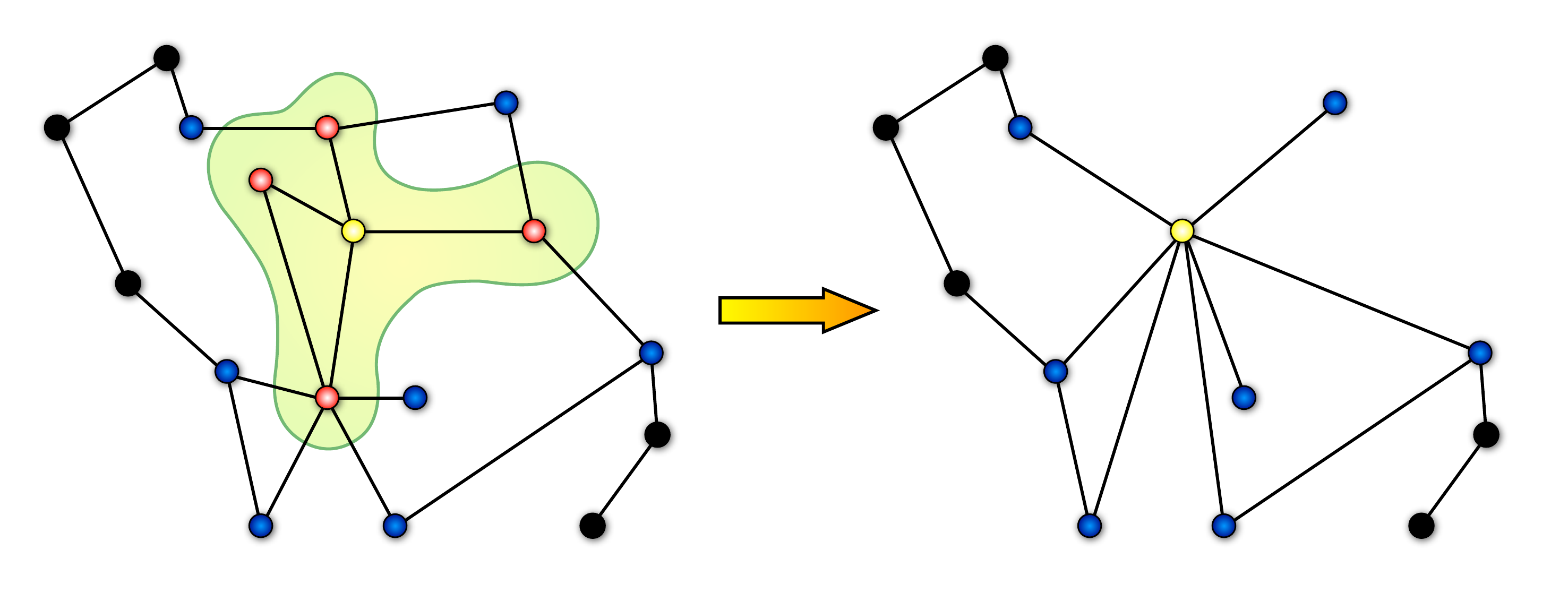}
\caption{(color online) One step of RSR with radius $b=1$. The randomly chosen target node 
(yellow/white in print) absorbs all its nearest neighbors (red/light gray in print). All 
links to the absorbed nodes (blue/dark gray nodes) are then redirected to the target. 
Alternatively one can view the supernode as a cluster that grows by eating all its neighboring clusters.
RSR with any $b>1$ can be performed by applying the above procedure on the same target $b$ times.    }
\label{fig:renorm}
\end{figure}

Random sequential renormalization, RSR, with radius $b~(b=1, 2, ... )$ is the process of 
consecutively applying a local coarse-graining (one step of RSR), on a given network $G_0$, 
which leads to a series of graphs $G_t (0 \leq t \leq T) $ with strictly decreasing sizes $N_t$. In the 
following, $t$ is called {\it time}. The initial graph at $t=0$ has $N_0$ nodes, and the final 
graph at $t=T$ is just a single node. We also assign masses to the nodes (initially $m_i=1; 
\forall i\in G_0$).  For any undirected and unweighted graph with $N$ nodes of masses ${m_i}$, 
one step of RSR (as shown in Fig.~\ref{fig:renorm}) proceeds as follows: \\
(i) Choose randomly one of the nodes in the graph as the target.\\
(ii) Define the neighborhood $\mathcal N$ around the target to include all nodes within a 
distance $d \leq b$ from the target. Distance is measured by the number of links that make the 
shortest path between two nodes. \\
(iii) Delete all the nodes in $\cal N$, except for the target.\\
(iv) Delete all the internal links of $\mathcal N$.\\
(v) Redirect to the target all links that connect nodes in $\mathcal N$ to the rest of the network.\\
(v) If a multiple link appears, replace it by a single link.\\
(vi) Update the mass of the target to $m=\sum_i m_i$, with $i\in \cal N$.\\

Hence, the target node and all its neighbors up to distance $b$ in the network are replaced by 
a super-node. This process preserves all the links to the outside but discards the internal 
details in the target's neighborhood --  analog to course-graining in real space renormalization. 
The super-node is then treated like any other node in the network. We consecutively repeat this procedure until the graph is reduced to a single node. 
Alternatively, one can also define RSR such that the target node is chosen with probability 
proportional to its mass~\cite{claireAP, swmassdep} or degree, but we only discuss the unweighted form here. 

For $b=1$ the target absorbs only its nearest neighbors. The easiest way to implement RSR with 
any $b>1$ is to apply RSR with $b=1$ on the same target for $b$ successive steps. Although 
this is slightly slower than an optimal coding, we use it in our simulations to reduce code 
complexity and potential sources of errors.

As indicated in Fig.~\ref{fig:renorm} RSR can also be interpreted as a cluster growth process on the graph. The target cluster is chosen at random and grows by absorbing all clusters within distance $b$ of it. Hence the fast Newman-Ziff (NZ) algorithm for growth of percolation clusters can be easily adapted to this problem, and it makes sense to speak of a percolation transition beyond which one of the clusters occupies a finite fraction of the nodes.  

\subsection{The graph ensembles under discussion}

We mainly focus on connected Erd\"os R\'enyi (ER) graphs with average degree $\langle k \rangle$ 
slightly larger than 2. The ensemble is produced in the following way: For each graph size $N_0$ 
we make several ER graphs with fixed size $N^* > N_0$ and a fixed number of links such that the 
average degree $\langle k\rangle^*=2$, and determine their giant component (which contains about 
80\% of the nodes for this value of $\langle k\rangle^*$). If the size of the giant component is 
$N_0 \pm 1\%$ -- corresponding to $N_0 = (0.80\pm 0.01)N^*$ -- we add the giant component to the ensemble; otherwise it is 
discarded. Notice that this leads to a slight scatter of $N_0$ and of the average degree of the 
graphs at the start of RSR. The latter is $\approx \langle k \rangle_0=2.4$. 
For each $N^*$ the ensembles typically contain $\approx 10^4$ networks, and we apply several 
realizations of RSR on each of them.

We also examine RSR on ER graphs with $\langle k\rangle^* \neq 2$, as well as the scale-free model 
of Barab\'asi and Albert~\cite{BA}. In each case the ensemble of connected graphs is generated in a similar manner.

\subsection{Algorithms and quantities of interest}

In network studies much attention has focused on the statistics of the number of links (degree) of nodes in a 
network. The degree distribution, the maximum degree, the average, and higher moments of 
the distribution are also often considered. 
But keeping track of the degrees of all nodes under RSR is time consuming and seriously 
confines the system sizes and statistics of numerical studies. In this paper we have 
performed numerical simulations with degree measurements for networks up to $N^*=2.4\times 10^5$
nodes. 

As mentioned previously, the NZ algorithm can be adapted to keep track of cluster masses rather than their degrees. With the NZ algorithm large network sizes 
with high statistics can be simulated in a reasonable time. We have performed RSR with mass 
analyses on networks up to $N^*=10^7$ nodes. Unfortunately all our efforts to track 
the degrees of the nodes using the NZ or other algorithms have led to extremely long running 
times, thus we restrict our analysis to degrees of  smaller networks and measure only masses
for larger ones.
As far as critical behavior is concerned, we show
that mass and degree distributions lead to similar conclusions.

\subsection{Averaging over the ensemble}

\begin{figure}
\includegraphics[width=1\columnwidth]{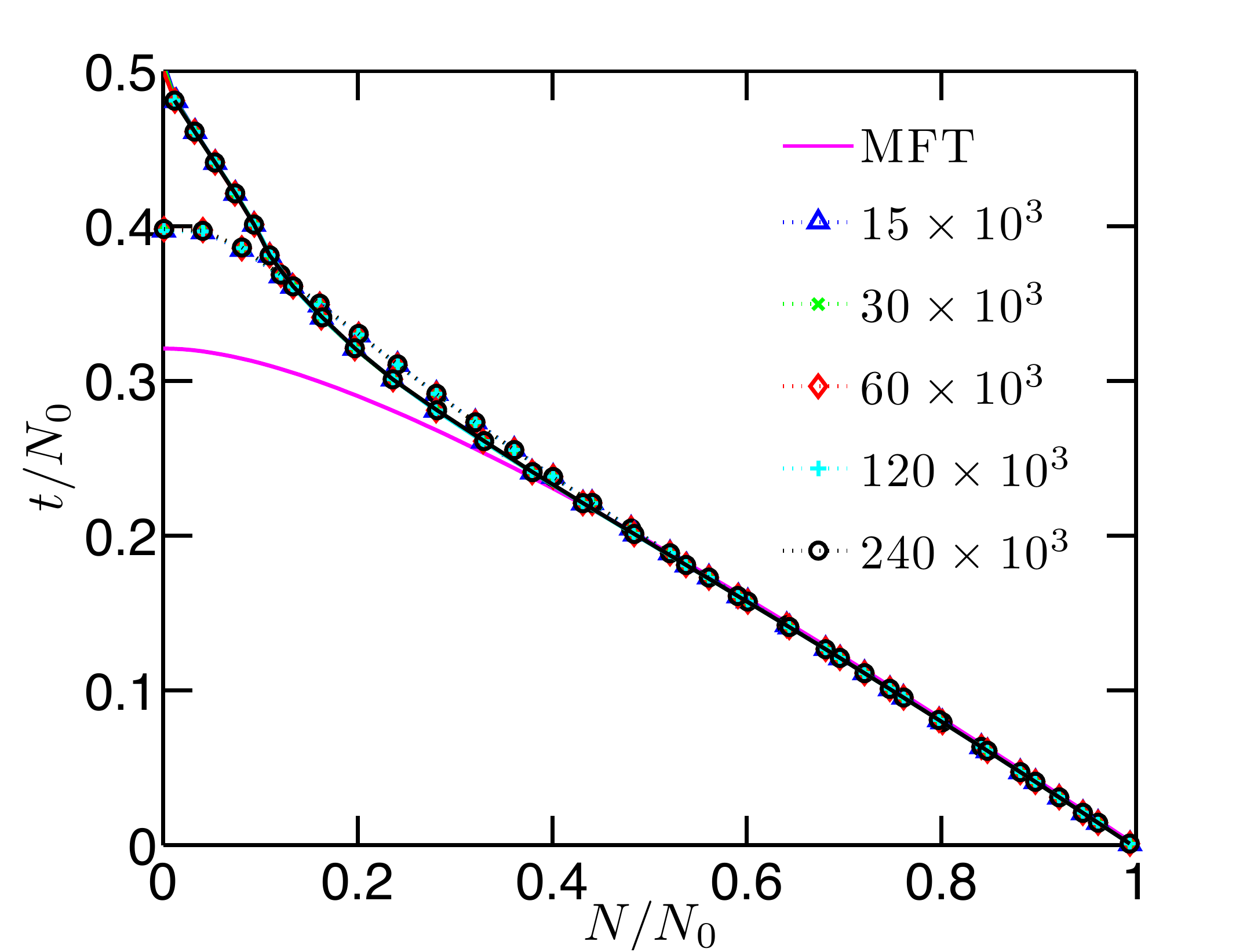}
\caption{(color online) Time dependence of network size, N, in rescaled units. The size decreases 
   monotonically under RSR. Data is obtained from ER graphs with $\langle k \rangle^*=2$ 
   under RSR with $b=1$. The curves with dashed lines are obtained by averaging $t$ values 
   corresponding to fixed $N$, and the curves with solid lines are obtained by averaging $N$ 
   for fixed $t$. The magenta (gray in print) solid line shows the mean-field theory 
   prediction (see Sec. IV). The two averages differ when mean-field theory breaks down due to 
   fluctuations. In the rest of the paper we choose $N$ as the independent variable and 
   average all other quantities at fixed $N$. Numbers in the legend show the initial size of 
   the ER graph, $N^*$, from which the initial giant components are obtained. }
\label{fig:Nt}
\end{figure}

When discussing ensemble averages, one can use different quantities as independent control 
parameters. In particular, one can average over RSR trajectories at fixed $N$ or at fixed $t$.
As shown in Fig.~\ref{fig:Nt}, these two ways of averaging give different results at late 
times (and hence small $N$), due to large fluctuations in the number of nodes eliminated 
per RSR step in the hub dominant phase. In the same figure we also show the result of a 
mean-field theory (MFT) discussed in section IV. During the initial stages of the flow, MFT
gives an accurate description of RSR, but breaks down when different ensembles lead to 
different results. Some RSR flows last much longer time than others and since we want to 
keep the number of members in the ensemble more or less fixed to obtain each data point, we 
choose to average at fixed $N$ (rather than $t$) in the rest of this paper.


\section{Simulation results for ER graphs with initial ${\bf \langle k \rangle^*=2}$}

We focus in detail on the behavior of the giant component of ER graphs with 
$\langle k \rangle^*=2$ under RSR with $b=1$. For these graphs the average degree of the 
giant component is $\langle k \rangle_0=2.4$. We find evidence for a continuous 
`agglomerative percolation' transition in the evolution of these networks under RSR. The 
transition is associated with the emergence of a giant hub or the percolation of a giant cluster on the network. 
We study scaling properties at this transition and measure the corresponding critical 
exponents numerically. We show that these exponents obey scaling relations associated with 
percolation, although RSR represents a different universality class than ordinary percolation, even in the mean-field 
limit.

\subsection{Evidence for a phase transition}

\begin{figure}
\includegraphics[width=1\columnwidth]{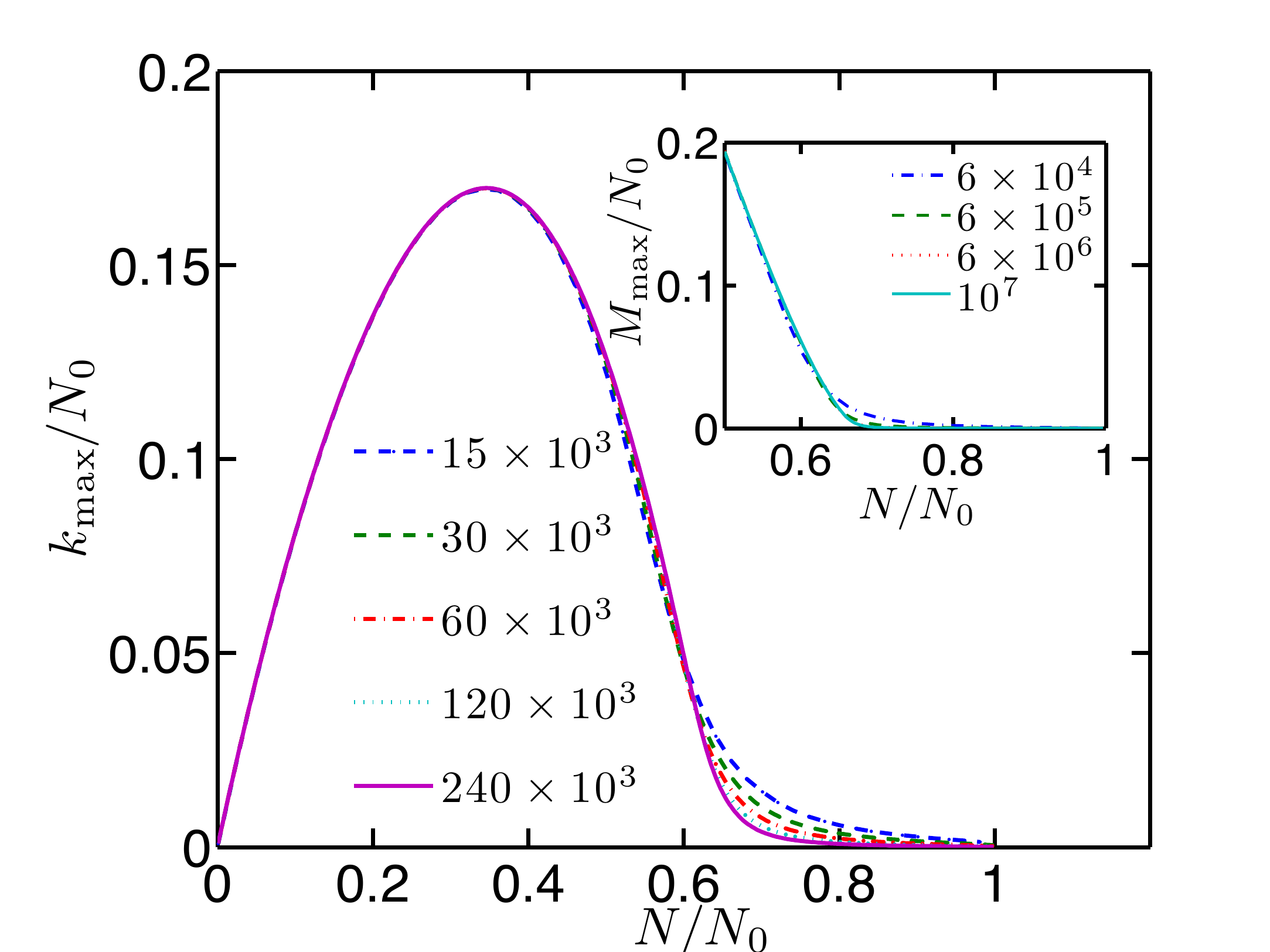}
\caption{(color online) Plot of $k_{\rm max}/N_0$ {\it vs.} $N/N_0$ for ER graphs with 
   $\langle k \rangle^*=2$ and several initial sizes. Note that the direction of the RSR 
   flow is from right to left. While $k_{\rm max}/N_0$ is close to zero in the mean-field 
   regime, the hub at late times absorbs a finite and increasing fraction of the nodes. The 
   transition gets sharper with increased system size; Inset: similar behavior for the 
   rescaled maximal cluster mass $M_{\rm max}/N_0$. Note that $M_{\rm max}$ always 
   increases monotonically under RSR, whereas $k_{\rm max}$ has to finally  decrease.  Using 
   the Newman-Ziff algorithm mass related properties can be measured
   on much larger systems than degree related properties.}
\label{fig:kmax}
\end{figure}

\begin{figure}
\includegraphics[width=1\columnwidth]{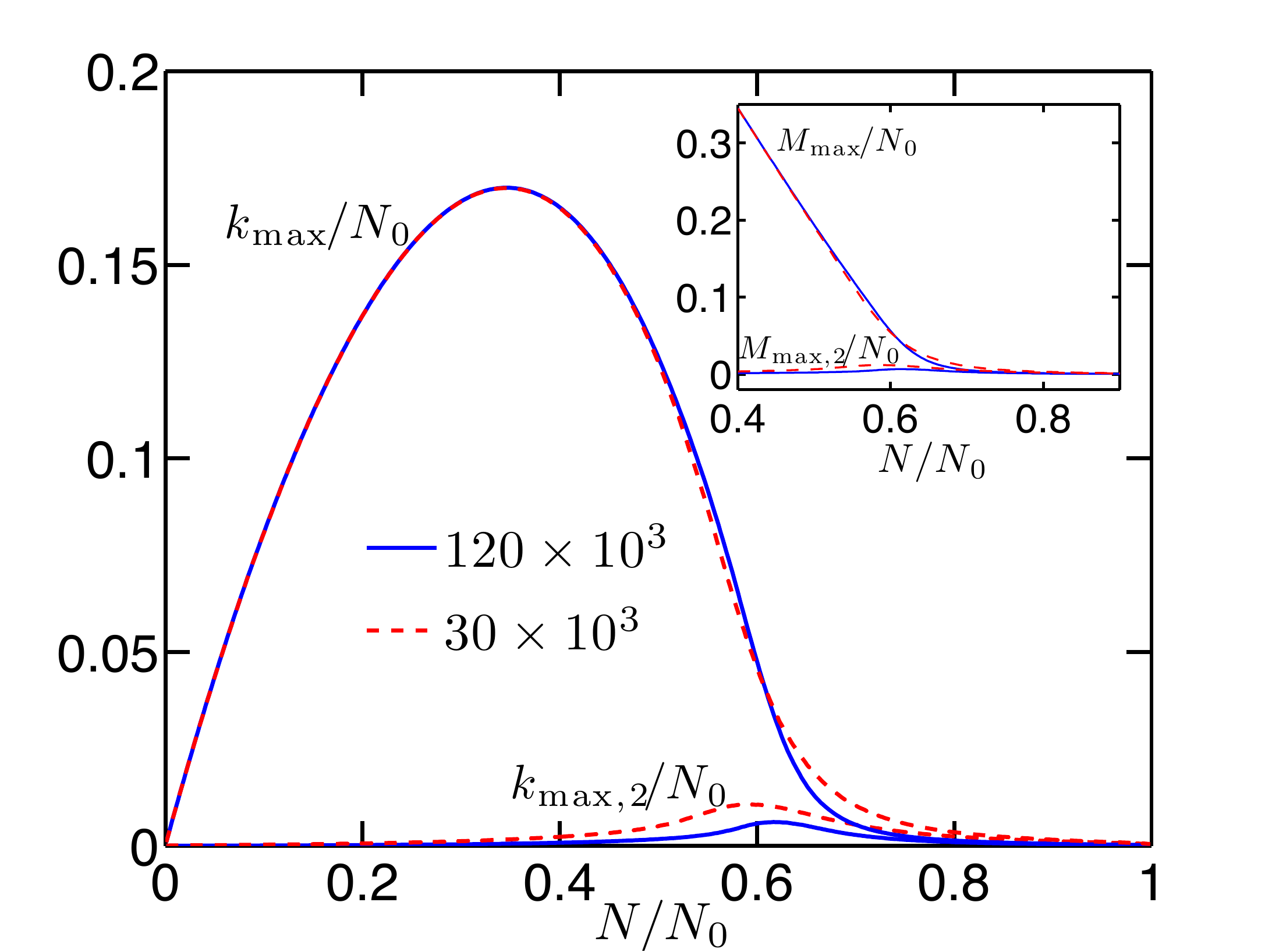}
\caption{(color online) The largest degree $k_{\rm max}$ and the second largest degree, 
   $k_{\rm max,2}$, are of comparable size in the mean-field regime, but in the hub regime a 
   giant hub takes over and the second largest degree shrinks. This behavior is also 
   consistent with a continuous percolation transition, and shows that there is only one 
   outstanding hub (cluster) in every network. The inset shows the same behavior for the 
   largest and the second largest mass. The data is obtained from ER graphs with $\langle k \rangle^*=2$.}
\label{fig:ratio}
\end{figure}

We begin by studying the behavior of the maximum degree $k_{\rm max}$, as a function of $N/N_0$, 
for various initial system sizes, as shown in Fig.~\ref{fig:kmax}. Note that the direction of the 
renormalization flow -- or {\it time}, is from right to left. The initial ER graph has a 
narrow Poisson degree distribution with no hubs and $k_{\rm max}/N_0$ is ${\cal O}(1/N_0)$. 
As RSR aggregates nodes locally, although higher degree nodes appear in the system, 
$k_{\rm max}/N_0$ remains small. However,  as shown in Fig.~\ref{fig:kmax}, 
$k_{\rm max}/N_0$ suddenly at $N/N_0\sim 0.7$, starts to increase more rapidly. This implies the existence of at least 
two regimes in the evolution of ER graphs under RSR: First, a {\it no-hub} (or mean-field; see
Sec.~IV) regime, where the degree distribution is narrow, fluctuations are negligible, and a mean-field theory describes the evolution of the system; Second, a {\it hub} regime 
where a growing hub exists and our mean-field theory breaks down. This 
is due to large fluctuations as well as the effect of loops in the network. Loops are 
present in the networks initially, but they typically are large and the graphs
are locally tree-like. As RSR proceeds, these loops become shorter and the graphs
no longer remain locally tree-like.
As indicated in Fig.~\ref{fig:kmax}, the transition between these two regimes becomes sharper on increasing system size $N_0$.

The same behavior can be observed for the mass of the larger cluster, $M_{\rm max}$, as shown in the inset 
of Fig.~\ref{fig:kmax}. Initially $m=1$ for all nodes. Although clusters grow under the 
renormalization flow in the mean-field regime, the maximum mass remains ${\cal O}(1)$. In 
the critical region a node with the largest mass percolates and separates itself from the rest 
of the distribution both in terms of size and degree. 

This is also indicated in Fig.~\ref{fig:ratio}, where both $k_{\rm max}$ and the second 
largest degree $k_{\rm max,2}$ are  plotted {\it vs.}  $N/N_0$. 
While the two largest degrees are about the same size in the mean-field regime, after 
the transition the largest hub grows and the second largest degree shrinks, which is 
another indication of a percolation transition. Similar behavior for $M_{\rm max}$ and 
the second largest mass, $M_{\rm max,2}$, is shown in the inset.

The detailed relation between mass and degree is discussed in the Appendix. No 
singular behavior in $k_{\rm max}~vs.~M_{\rm max}$ appears in the critical region, and 
this smoothness holds statistically for the mass and degree of other nodes as well. Thus either variable can 
be used to extract the critical properties of the phase transition. Since RSR with mass 
measurement is much faster using the NZ algorithm, we mostly base our discussions on 
the masses of nodes.  

\subsection{Finite size scaling analysis}

\begin{figure}
\includegraphics[width=1\columnwidth]{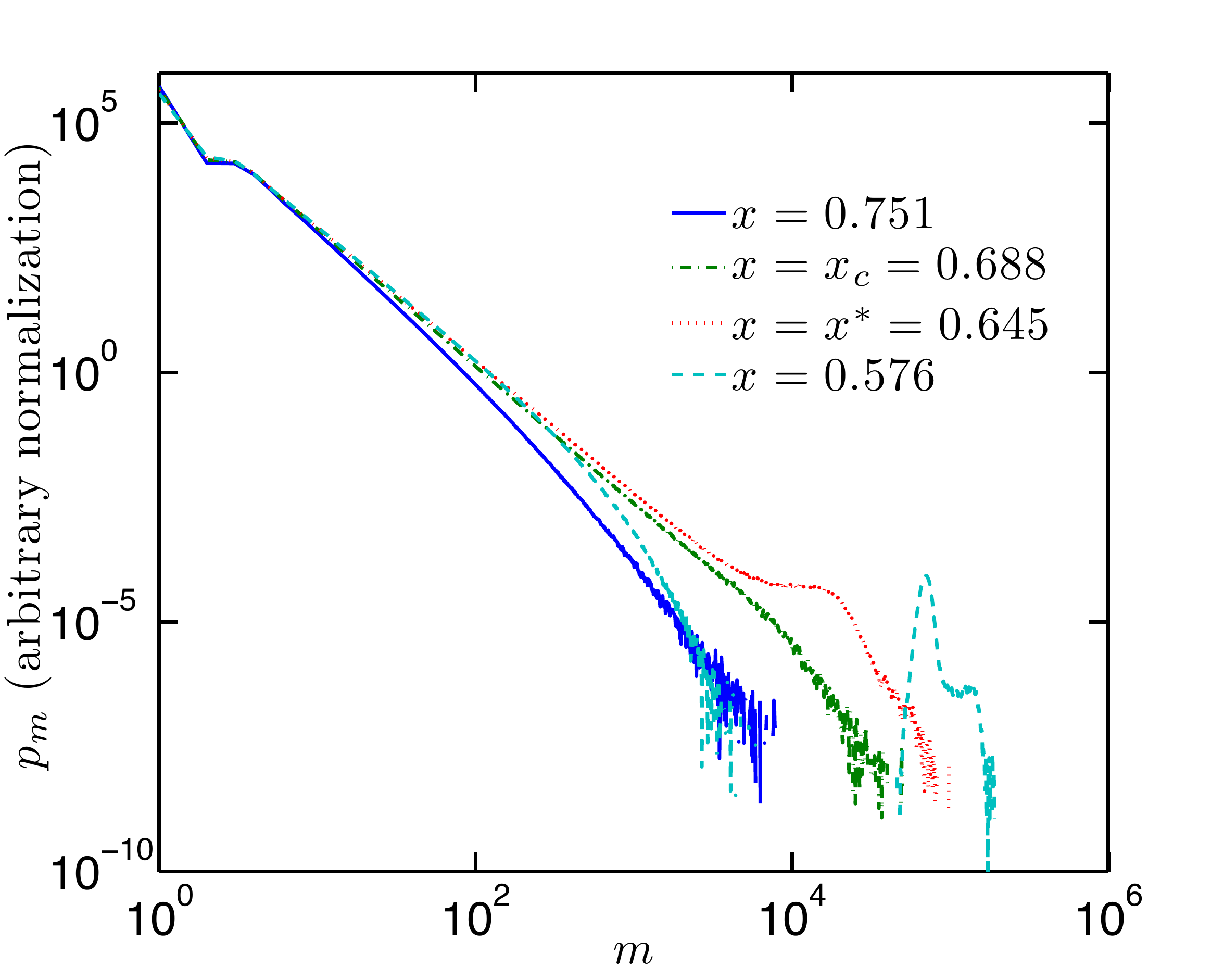}
\caption{(color online) Cluster mass distribution at different stages of the RSR flow for ER 
   graphs of $N^*=10^6$ nodes, and $\langle k\rangle^* =2$. This distribution broadens and approaches a power-law 
   $p_m\sim m^{-\tau}$ as $x=N/N_0$ decreases. The power-law is broadest at $x^*(N_0)=0.645$, 
   for this system size. For $N_0\to \infty$, the critical point converges to $x^*\to x_c=0.688$ 
   (the green / dashed-dotted curve). For $x<x^*$ a giant cluster emerges and a gap expands 
   between this cluster and the rest of the distribution. Note that the size distribution of 
   the giant cluster has a shoulder on the right (unlike ordinary percolation). This is due 
   to the possibility of selecting the hub as a target node and is discussed 
   in more detail in Sec.~V.}
\label{fig:massdist}
\end{figure}

\begin{figure}
\includegraphics[width=1\columnwidth]{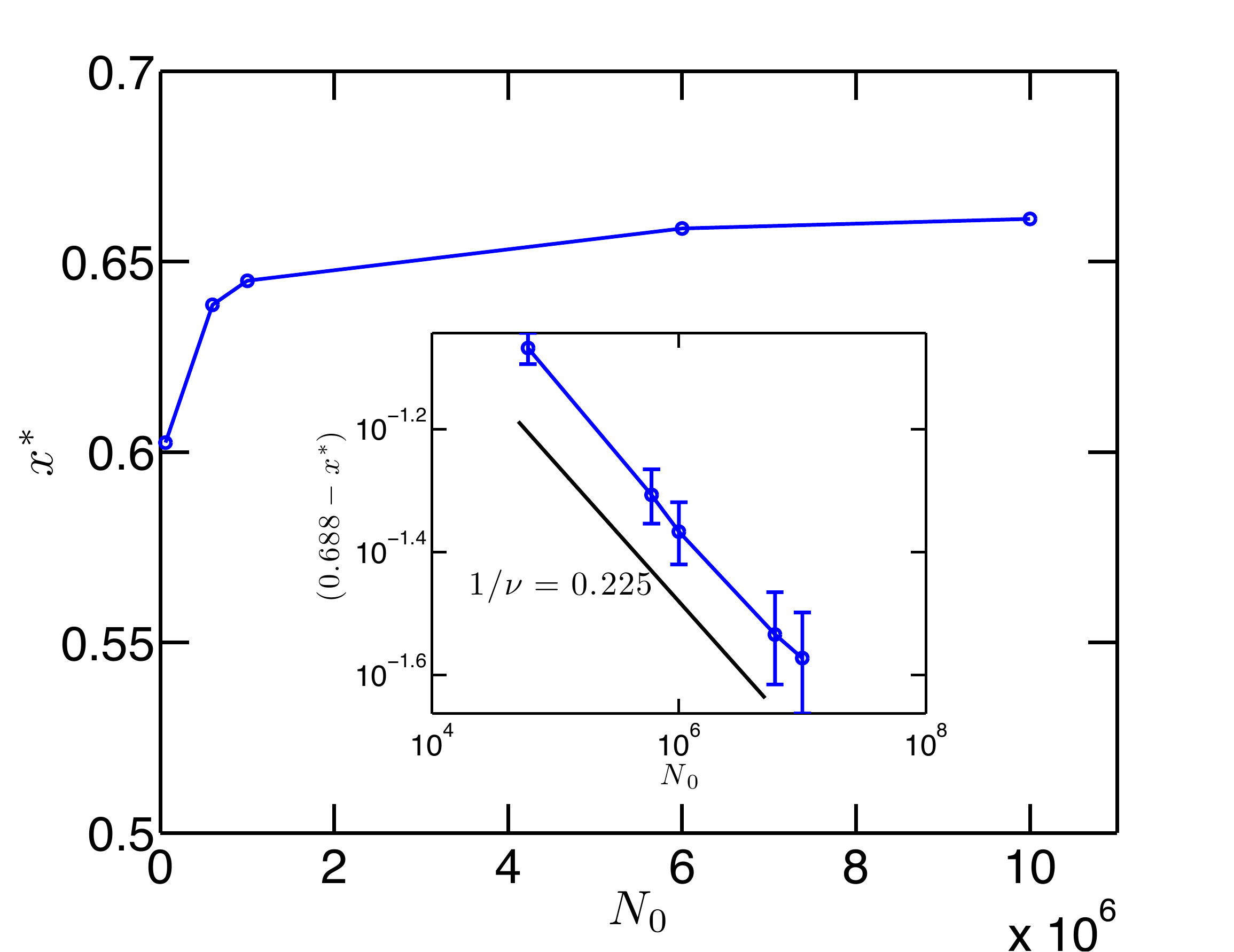}
\caption{(color online) Convergence of the effective critical point, $x^*(N_0)$, to $x_c=0.688$ 
   as the system size increases. Inset: the critical point $x_c$ and the exponent $1/\nu$ are 
   consistent with the values $x_c=0.688$ and $-1/\nu=-0.225$, as indicated by the slope of the straight line.} 
\label{fig:xc-N0}
\end{figure}

\begin{figure}
\includegraphics[width=1\columnwidth]{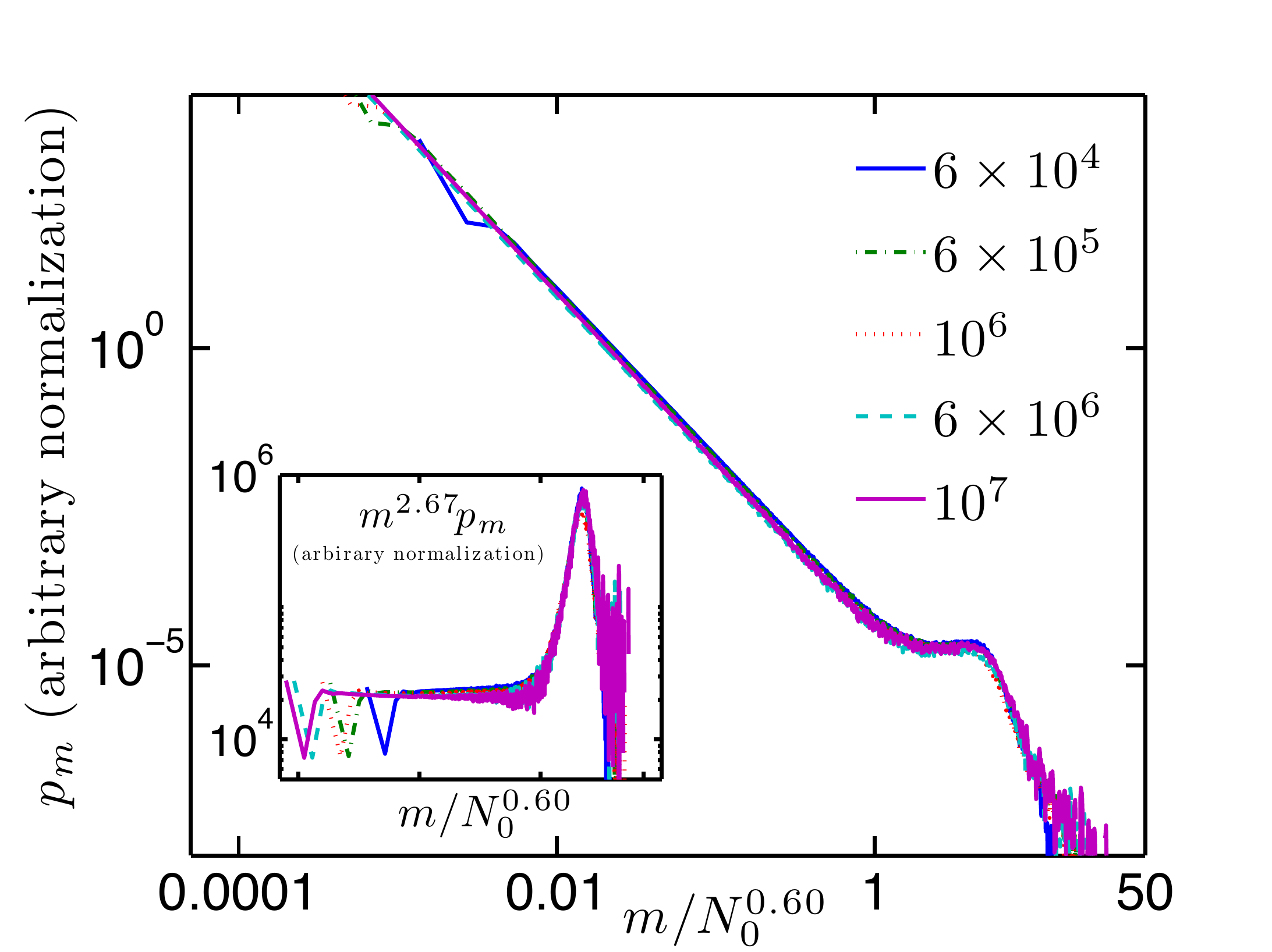}
\caption{(color online) Data collapse using the FSS ansatz in Eq.~(\ref{eq:pm-scale}) for the 
   mass distribution at $x^*(N_0)$. The exponent $D=0.6$ gives the best data collapse, and 
   $\tau=2.67$ fits the power law (see the inset). These values are consistent with Eq.~(\ref{eq:tau}).}
\label{fig:massdist-collapse}
\end{figure}

In order to analyze the RG flow in the critical region, we perform a finite size scaling (FSS) 
analysis on a number of quantities and their distributions. Initially all the nodes have mass $m=1$. 
As shown in Fig.~\ref{fig:massdist}, the mass distribution broadens with the number of RSR steps, 
until a power law distribution $p_m\sim m^{-\tau}$ emerges in the critical region. 
As the RSR flow continues, an expanding gap appears between the giant cluster and the rest of 
the clusters. As shown by the curve for $x\equiv N/N_0=0.576$ (below the transition), the peak 
corresponding to the giant cluster has a pronounced shoulder on the right. This is different from 
ordinary percolation (OP), where the peak is featureless. As
discussed in more detail in Sec.~V, this shoulder results from the giant cluster being  chosen repeatedly as the target of RSR. These are rare events, but they have dramatic effects on the flow.

Setting  $x=N/N_0$, the effective critical point for a finite system, $x^*(N_0)$, is defined as 
the value at which the system has the broadest power-law in its mass distribution. In Fig.~\ref{fig:xc-N0}, 
we illustrate the convergence of $x^*$ as the system size increases. The limiting value for infinite 
system size, $x_c$, is consistent with $x_c=0.688$ as shown in the inset.

To proceed further, we make a conventional scaling ansatz for the mass distribution of a finite system
in terms of a homogeneous scaling function~\cite{Stauffer}
\be
   p_m=m^{-\tau} g(nN_0^{1/\nu}, m/N_0^D) \;\;,
\label{eq:pm-scale}
\ee
where 
\be
   n=(x-x_c)/x_c \;\;.
\ee
Note that such an ansatz is never perfect, and all critical parameters discussed in the following are 
obtained by compromises to get the best overall data collapses for all quantities of interest, and by
assuming the scaling relations between critical exponents implied by the FSS ansatz. A summary of all 
critical exponents, the equations defining them and the figures demonstrating numerical evidence, are 
given in Table~\ref{T:1}. 

Our best estimate for the critical point -- mainly from Fig.~\ref{fig:xc-N0}, but also taking into 
account the consistency checks in subsection C -- is 
\be
x_c=0.688 \pm 0.002 \;\;.
\ee
The exponent $1/\nu$ in Eq.~(\ref{eq:pm-scale}), describing the convergence of  $x^*$ to $x_c$ with 
increase of system size, is determined to be 
\be
1/\nu=0.225 \pm 0.005 \;\;.
\ee
The exponent $D$, giving the scaling of the maximum mass with system size (see 
Fig.~\ref{fig:massdist-collapse}), is 
\be
D=0.60 \pm 0.01\;\;.
\ee
It is related to the Fisher exponent $\tau$ by demanding that there is $O(1)$ cluster of size 
$\geq N_0^D$ and using Eq.~(\ref{eq:pm-scale})~\cite{Stauffer}:
\be
   \tau={1+D\over D}=2.67\pm0.03 \;\;.     \label{eq:tau}
\ee


\begin{figure}
\includegraphics[width=1\columnwidth]{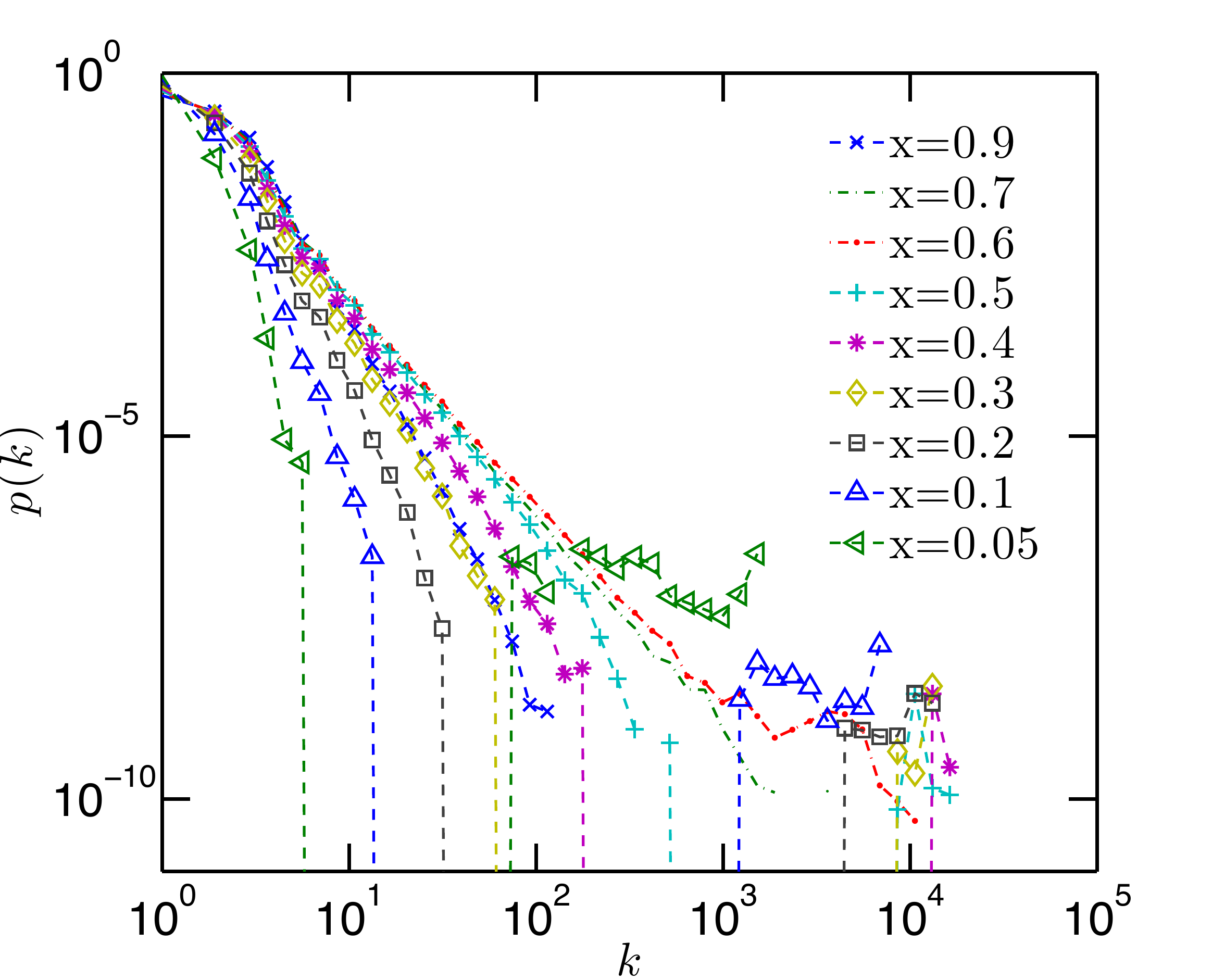}
\caption{(color online) Degree distributions for $N^*=1.2\times 10^5$ at different 
   stages of the RSR flow. The initial, narrow distribution gets broader and approaches 
   a power law $p_k=k^{-\tau_k}$ close to the transition. Then a giant hub stands 
   out and a gap opens between the hub and the rest of the nodes. }
\label{fig:degdist}
\end{figure}

Degree distributions behave similar to the mass distributions. The initial ER graph 
has a Poisson degree distribution. As RSR proceeds, higher degree nodes appear and 
the degree distribution broadens. At the phase transition local hubs join together 
to make a single hub much larger than all others. Just before the giant hub emerges, 
the degree distribution is approximately a power law with a power $\tau_k$ that is 
consistent with the power $\tau$ of the mass distribution. Afterward, the hub continues 
to grow, but not forever. Figure~\ref{fig:degdist} shows the degree distribution at 
different values of $x=N/N_0$. At $x\approx 0.7$ the distribution resembles a power law, at 
$x\approx0.6$ a bump appears at the rightmost end of the distribution. Later, when 
$x<0.6$, a growing gap forms between the giant hub and the rest of the nodes. This 
continues until the shrinking system size forces the degree of the giant hub to decrease.


\subsection{Consistency checks}

\begin{table*}[t]
\begin{center} 
\begin{tabular}{| c | l | l | c | c | c | c|} 
\hline \hline
  exponent & ~ value (RSR)    & scaling relation & Eq(s).                             & Fig(s).                              & mean-field OP & mean-field AP \\ \hline
     $D$   & $0.60 \pm 0.01$  &		         &\ref{eq:pm-scale}                   & \ref{fig:massdist-collapse}                  & $2/3$ &   --   \\ 
    $\nu$  & $4.44\pm0.10$     &                  &\ref{eq:pm-scale}                   &\ref{fig:xc-N0}, \ref{fig:massdist-collapse}  & $3$   &   $4.4\pm0.3$   \\
   $\tau$  & $2.67\pm0.03$    & ~$(1+D)/ D$      &\ref{eq:pm-scale}, \ref{eq:tau}     &\ref{fig:massdist-collapse}                   & $5/2$ &   3    \\
  $\beta$  & $1.78\pm0.08$    & ~$(1-D)\nu$      &\ref{eq:Mmax}, \ref{eq:s1},  \ref{eq:Mmax1}      & \ref{fig:mmax_scale}           & $1$   & -- \\
 $\sigma$  & $0.375\pm0.015$  & ~$1/(D\nu)$      &\ref{eq:sig}, \ref{eq:sig2}, \ref{eq:sigma-scale} &   \ref{fig:max2}               & $1/2$ &   --   \\
 $\gamma$  & $0.88\pm0.10$     & ~$2D\nu-\nu$     &\ref{eq:gam}, \ref{eq:gam1}, \ref{eq:gamma-scale} & \ref{fig:M2er_scale}           & $1$   &   1/2  \\
 $\alpha$  & ~$6.8 \pm 0.3$   & ~see text        &\ref{eq:alpha}                      & \ref{fig:chi_k}                              & $4$   &   --   \\
\hline 
\end{tabular} 
\caption{Summary of critical exponents for ER graphs with $\langle k\rangle^*=2$ under 
  RSR with $b=1$. All exponents are obtained by best compromise for 
  the data collapses shown in the figures listed in column 5, and by requiring the 
  scaling relations in column 3 to hold, except for the exponent $\alpha$ (last line). 
  The critical exponents are clearly different
  from those of mean-field ordinary percolation (column 6) and for mean-field agglomerative 
  percolation(last column).}
\label{T:1}
\end{center}
\end{table*}

\begin{figure}
\includegraphics[width=1\columnwidth]{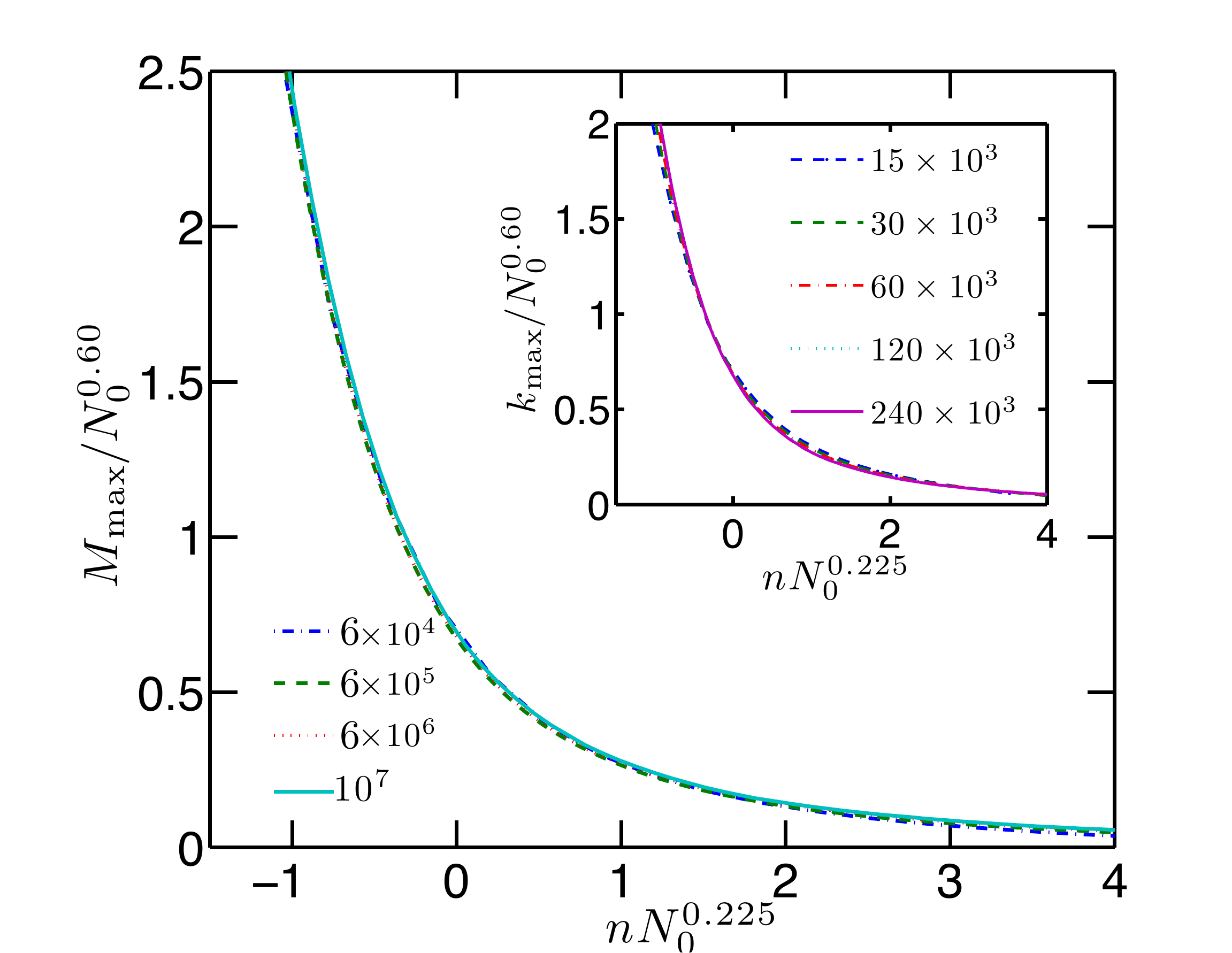}
\caption{(color online) Scaling plot of $M_{\rm max}$ in the critical region. The value of $x_c=0.688$
   in here and the following figures is the same as in Fig.~\ref{fig:xc-N0}, and the critical 
   exponents are those given in Table~I. 
   The inset shows that the exponents $\nu_k$ and $\beta_k$ for the maximum degree are the same 
   as those for the maximum mass within error.}
\label{fig:mmax_scale}
\end{figure}


In this subsection we check for consistency of our simulations with the scaling theory 
based on the FSS ansatz in Eq.~(\ref{eq:pm-scale}), by showing data collapses for 
different quantities of interest. Notice that the well known scaling relations between
critical exponents~\cite{Stauffer} follow from Eq.~(\ref{eq:pm-scale}) by considering 
appropriate limits.

\subsubsection{The order parameter}

An order parameter is any property of a system that can unravel the singularity at the 
critical point, which is non-zero only on one side of the transition. Typically $P_\infty$, the probability that a given site belongs to the 
percolating cluster, is considered as an order parameter for percolation. 
For RSR on graphs both $k_{\rm max}/N_0$ and $M_{\rm max}/N_0$, can be used as order 
parameters. Notice that the latter is equal to $P_\infty$.

An FSS ansatz for $M_{\rm max}$ follows by multiplying Eq.~(\ref{eq:pm-scale}) by $m^y$,
integrating over $m$, and taking the limit $y\to\infty$. Using also Eq.~(\ref{eq:tau})
gives 
\be 
   {M_{\rm max}\over N_0^D} = h(nN_0^{1/\nu})\;.     \label{eq:Mmax}
\ee
Assume now that $h(z)$ satisfies a power law, $h(z) \sim z^\beta$ for $z\to 0$. 
Equation~(\ref{eq:Mmax}) gives then in the supercritical case $n<0$ (where we expect 
$M_{\rm max} \propto N_0$) 
\be
   D=1-{\beta \over \nu}                             \label{eq:s1}
\ee
and
\be
   {M_{\rm max} \over N_0} \sim |n | ^{\beta}\;.                 \label{eq:Mmax1}
\ee

Figure~\ref{fig:mmax_scale} shows a data collapse according to Eq.~(\ref{eq:Mmax}), with 
$x_c=0.688$ and critical exponents as given in Table~I.
The analogous FSS ansatz for $k_{\rm max}$, 
\be
{k_{\rm max}\over N_0^{1-\beta_k/\nu_k}} = h_k(nN_0^{1/\nu_k})\;,
\ee
with $\nu_k=\nu$ and $\beta_k=\beta$ is shown in the inset of Fig.~\ref{fig:mmax_scale}.
The exponents for maximal mass and degree are equal within our errors.

\subsubsection{The cutoff scale for the cluster size distribution}

\begin{figure}
\includegraphics[width=1\columnwidth]{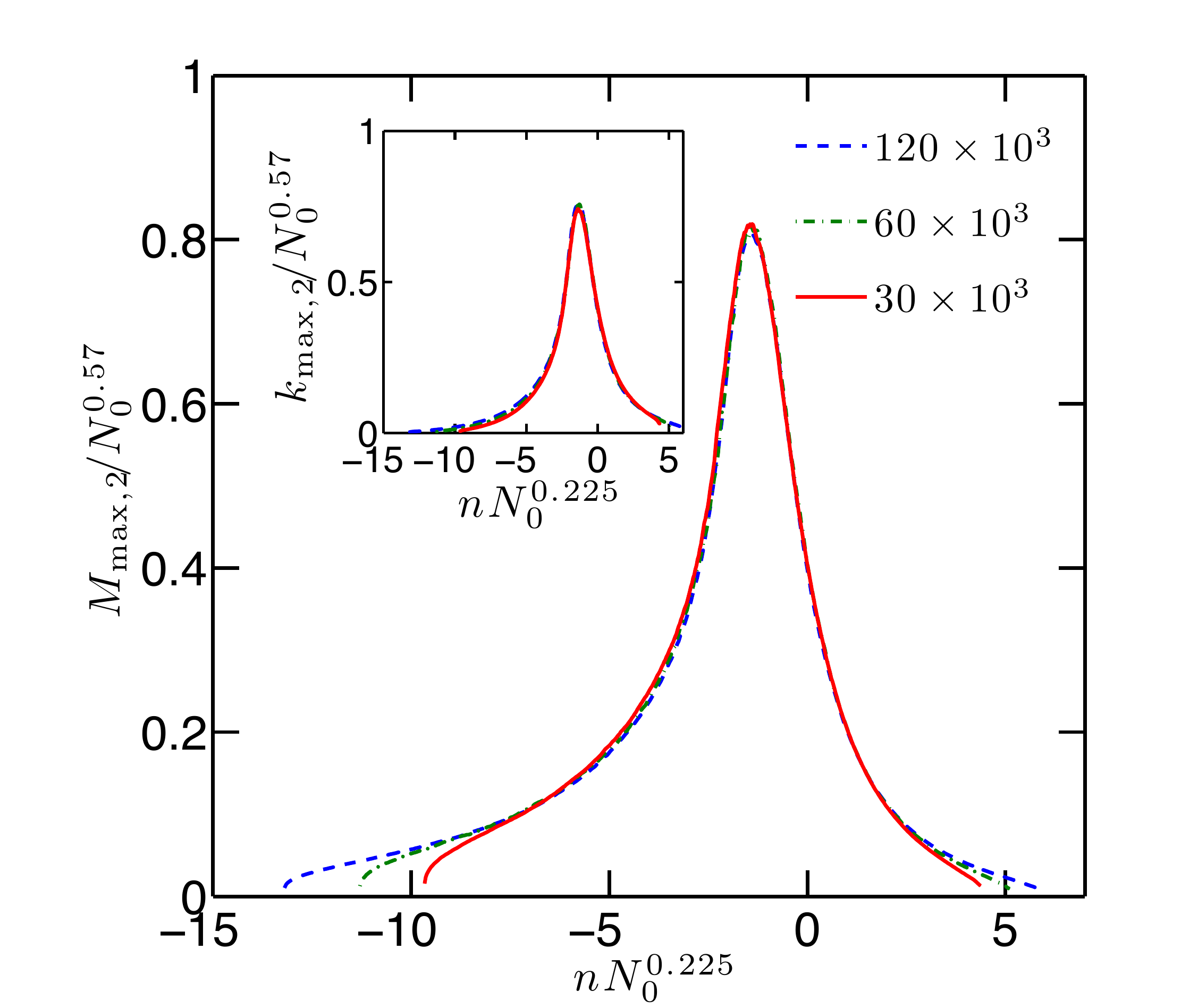}
\caption{(color online) Data collapse for the second largest mass and the second largest 
   degree. Again $\nu$ is taken from Table I, while $\sigma$ is fitted for optimal collapse.
   The value of $\sigma_k$ used in the inset is equal to $\sigma$, which is here $\sigma=0.395$.}
\label{fig:max2}
\end{figure} 

The size of the second large cluster, $M_{\rm max,2}$ (resp. $k_{\rm max,2}$), 
determines the cutoff for the finite clusters (excluding the hub). 
An FSS ansatz based on Eq.~(\ref{eq:pm-scale}) gives
\be 
\frac{M_{\rm max,2} }{N_0^{1/\sigma \nu}} = h_2(nN_0^{1/\nu})
\label{eq:sig}
\ee
and for the infinite system limit
\be 
M_{\rm max,2} \sim n^{-1/\sigma} \quad.
\label{eq:sig2}
\ee
The exponent $\sigma$ is related to other exponents by
\be 
\sigma=\frac{1}{D\nu}=0.375\pm 0.015 \quad.
\label{eq:sigma-scale}
\ee
One can write similar equations for $k_{\rm max,2}$.  Figure~\ref{fig:max2} shows data collapse 
plots with $1/\sigma \nu=1/\sigma_k \nu=0.57$ for the second largest mass and degree. This leads to
\be
 \sigma=\sigma_k=0.395 \quad .
\ee
This estimate was chosen as it gives the best data collapse. It is consistent with the value obtained 
in Eq.~(\ref{eq:sigma-scale}), within error.


\subsubsection{Average cluster size}


\begin{figure}
\includegraphics[width=1\columnwidth]{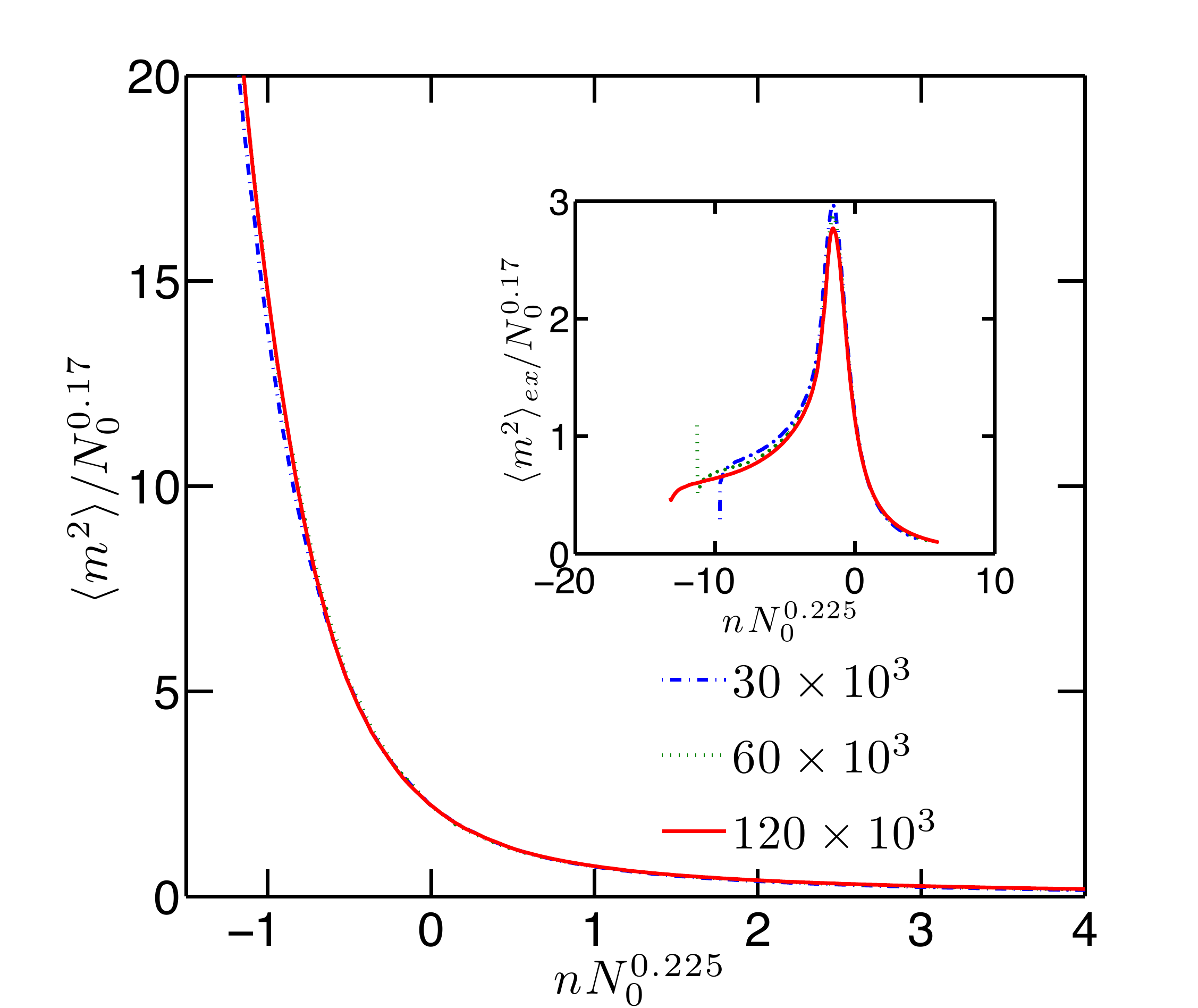}
\caption{(color online) Scaling plot of the second moment of the mass distribution,
   $\langle m^2\rangle$, for ER graphs, with $\gamma/\nu=0.17\pm 0.03$. The inset shows the 
   same plot for $\langle m^2\rangle _{ex}$. The same exponents are obtained.}
   \label{fig:M2er_scale}
\end{figure} 

\begin{figure}
\includegraphics[width=1\columnwidth]{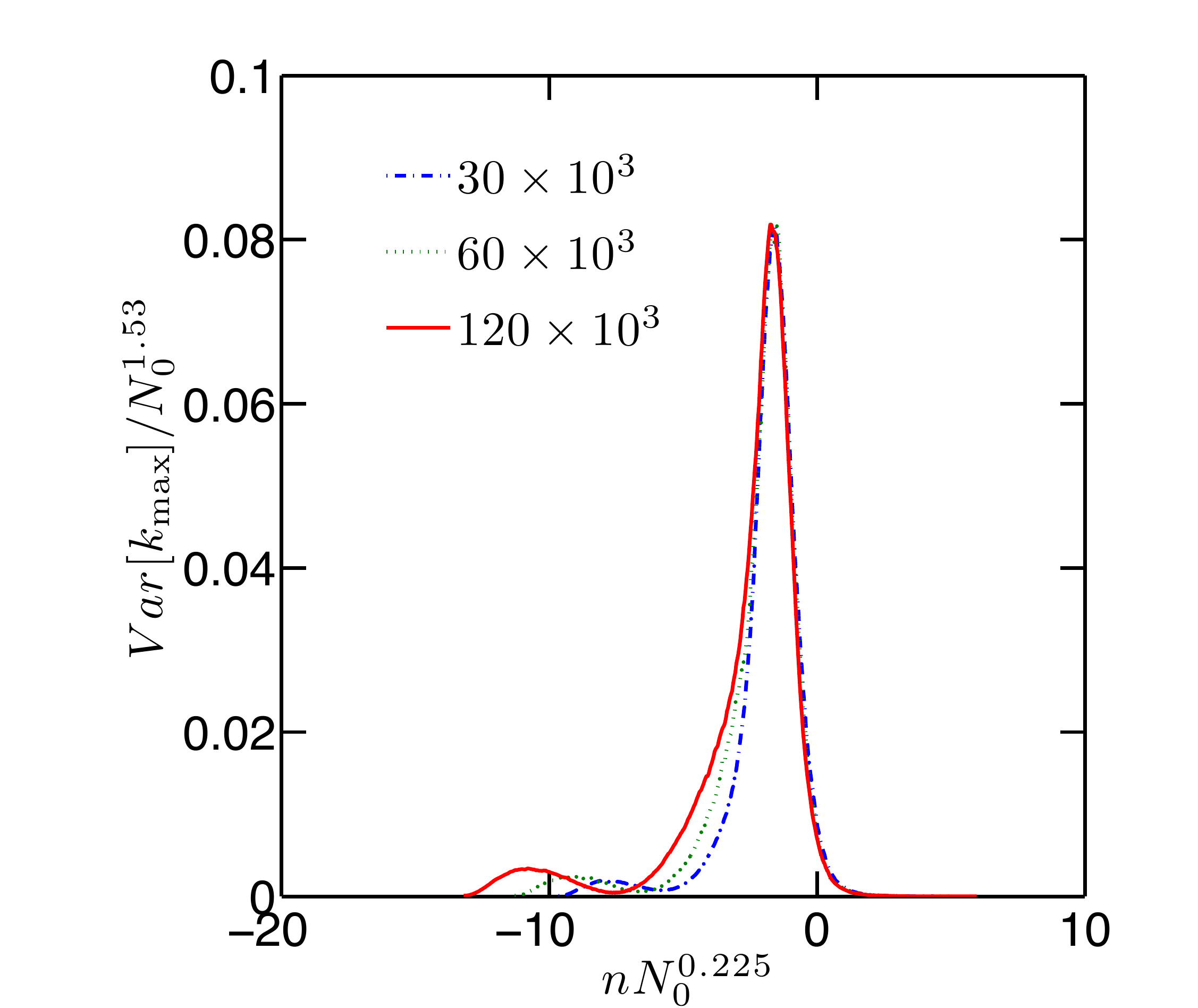}
\caption{(color online) Scaling plot for the variance of the maximum degree, 
   with $x_c=0.688$, $1/\nu=0.225$ and $\alpha_k/\nu=1.53\pm0.02$. }
   \label{fig:chi_k}
\end{figure}

The average size of the cluster to which a randomly chosen node of the original network belongs 
is equal to the second moment of the mass distribution. An FSS ansatz for the average cluster 
size can be written as
\be
{ \langle m^2 \rangle \over N_0^{\gamma/\nu}}=J(nN_0^{1/\nu}) \;,
\label{eq:gam}
\ee
and in the limit of $N_0\to \infty$
\be
\langle m^2\rangle =n^{-\gamma} \;.
\label{eq:gam1}
\ee
The exponent $\gamma$ obeys the scaling relation  
\be
\gamma=(2D-1)\nu=0.88  \pm 0.10\quad.
\label{eq:gamma-scale}
\ee
Figure~\ref{fig:M2er_scale} shows the corresponding FSS analysis, with $\gamma/\nu=0.17$ 
chosen for an optimal data collapse. Within errors, this is consistent with the value 
$\gamma/\nu=0.20\pm 0.20$ obtained in Eq.~(\ref{eq:gamma-scale}). In the inset we show the 
second moment of the mass distribution excluding the largest cluster, $\langle m^2 
\rangle_{ex}$, with the same scaling exponents. 
The exponent $\gamma_k$ for the degree moment is found to be the same as that for the mass moment within error (data not shown).


\subsubsection{Variance of the maximal cluster size}

The variance of $k_{\rm max}$ and $M_{\rm max}$ also diverge at the critical point. 
Because of technical problems we do not have precise values of the latter, and we 
concentrate on the variance of $k_{\rm max}$. It should scale as
\be
{Var[k_{\rm max}]\over N_0^{\alpha_k/\nu_k}}=J'(nN_0^{1/\nu_k}).
\label{eq:alpha}
\ee
Figure~\ref{fig:chi_k} shows the corresponding scaling plot. In OP the standard 
deviation of the order parameter has the same critical exponent as the order parameter 
itself~\cite{subcritical}, implying $\nu-\beta=\alpha/2$. This is not what we find 
for RSR on ER graphs, if we assume $\alpha = \alpha_k$. To clarify this we directly 
looked at the distribution of $k_{\rm max}$ at $x^*$ for each of the system sizes 
shown in Fig.~\ref{fig:chi_k} (data not shown). The distribution is flat on the left 
side, but has an approximate power-law tail on the right. The fluctuations in this 
case grow faster than the average, unlike in ordinary percolation. We thus believe
that the observed violation of the scaling relation is not due to $\alpha \neq \alpha_k$,
but shows that the relation $\nu-\beta=\alpha/2$ is violated in AP.

This stems from the difference in the growth process in agglomerative and ordinary
percolation. Adding a bond (or site) in OP might merge only a few clusters into 
the giant cluster, leading to an additive growth of its size (and degree). In contrast, if the hub 
is chosen as the RSR target, it absorbs all its neighboring clusters. This leads to 
multiplicative growth. Thus in RSR we expect to see larger relative fluctuations in the 
hub size near the transition comparing to OP (see also Sec. V).


\section{mean-field theory and an annealed model}

We now approach the problem analytically using a mean-field theory (MFT) based on 
generating functions~\cite{newman01}. We will show that the critical exponents for 
mean-field RSR do not agree with the ones for ordinary mean-field percolation.

\subsection{General formalism}

Let $n_k$ be the number of nodes with degree $k$. The total number of nodes in the network 
is $N=\sum_k n_k$ and the probability of picking a node with degree $k$ is $p_k=n_k/N$. 
The change of $n_k$ in one step of RSR can be written as the sum of  a loss term $r_k$ 
associated with eliminating a $k$-degree node and a gain term $q_k$ associated with 
creating one,
\be
\frac{d n_k}{dt}=r_k+ q_k \;\;.
\label{eq:nk}
\ee
The loss term $r_k$ is
\be
r_k=-p_k-\sum_{k'} k' {kp_k \over \sum_{l}l{p_l}}p_{k'} = -(k+1)p_k \;\; .
\label{eq:rk}
\ee
The first term in the central expression is the probability of targeting a $k$-degree node, and the 
second term is the probability that any of the neighbors of the target have degree $k$. 
Note that the mean-field assumption is to ignore any potential correlations between the 
degrees of neighboring nodes.

In order to obtain an equation for $dN/dt$ one does not need to know $q_k$ in 
detail; one just has to know that exactly one new node is created, whence $\sum_k q_k=1$.
Summing Eq.~(\ref{eq:nk}) over $k$ leads then indeed to 
\be
   {dN\over dt}=-\langle k \rangle \;\;,  \label{eq:Nt}
\ee
as expected from the fact that all neighbors of a randomly chosen node are eliminated
in one RSR step.

To get $q_k$, assume that the target has $m$ neighbors with degrees $k_1, k_2,...k_m$. 
The new degree of the target will be the number of its second nearest neighbors. If 
all degrees are uncorrelated and the target's neighbors are not connected among themselves,
\be
   q_k= \sum_m p_m\sum_{k_1, k_2,...k_m}\prod_{i=1}^{m} {k_i p_{k_i} \over \langle k 
         \rangle}\delta_{k_1+...+k_{m},k+m} \;\;.
   \label{eq:qk}
\ee

We use generating functions to proceed. The degree distribution is generated by 
\be
G(x)=\sum_k p_k x^k \;\;,
\label{eq:G}
\ee
and  $q_k$ by:
\be
Q(x)=\sum_k q_k x^k \;\;.
\ee
The degree distribution of the neighbors of the target is proportional to $k p_k$, 
thus their remaining degree is generated by
\be
{\sum_k k p_k x^{k-1} \over \sum_k k p_k}={G^{'}(x) \over \langle k \rangle} \;\;.
\ee
Equation (\ref{eq:qk}) gives then
\be
   Q(x)=\sum_m p_m \left ( {G^{'}(x)\over \langle k\rangle}\right )^m = G 
   \left ( {G^{'}(x)\over \langle k \rangle}\right) \;\;.
   \label{eq:Q}
\ee
Using Eqs.~(\ref{eq:nk}) through~(\ref{eq:Q})
one can write the master equation for the generating function of the degree distribution as
\be
   {d \over dt} G(x)= {1\over N} \left [G \left ( {G^{'}(x)\over \langle k \rangle} 
   \right ) + \langle k-1  \rangle G(x) - x G^{'}(x)\right ]  \;\;.
\label{eq:dGdt}
\ee

\subsection{The average degree}

\begin{figure}
\includegraphics[width=1\columnwidth]{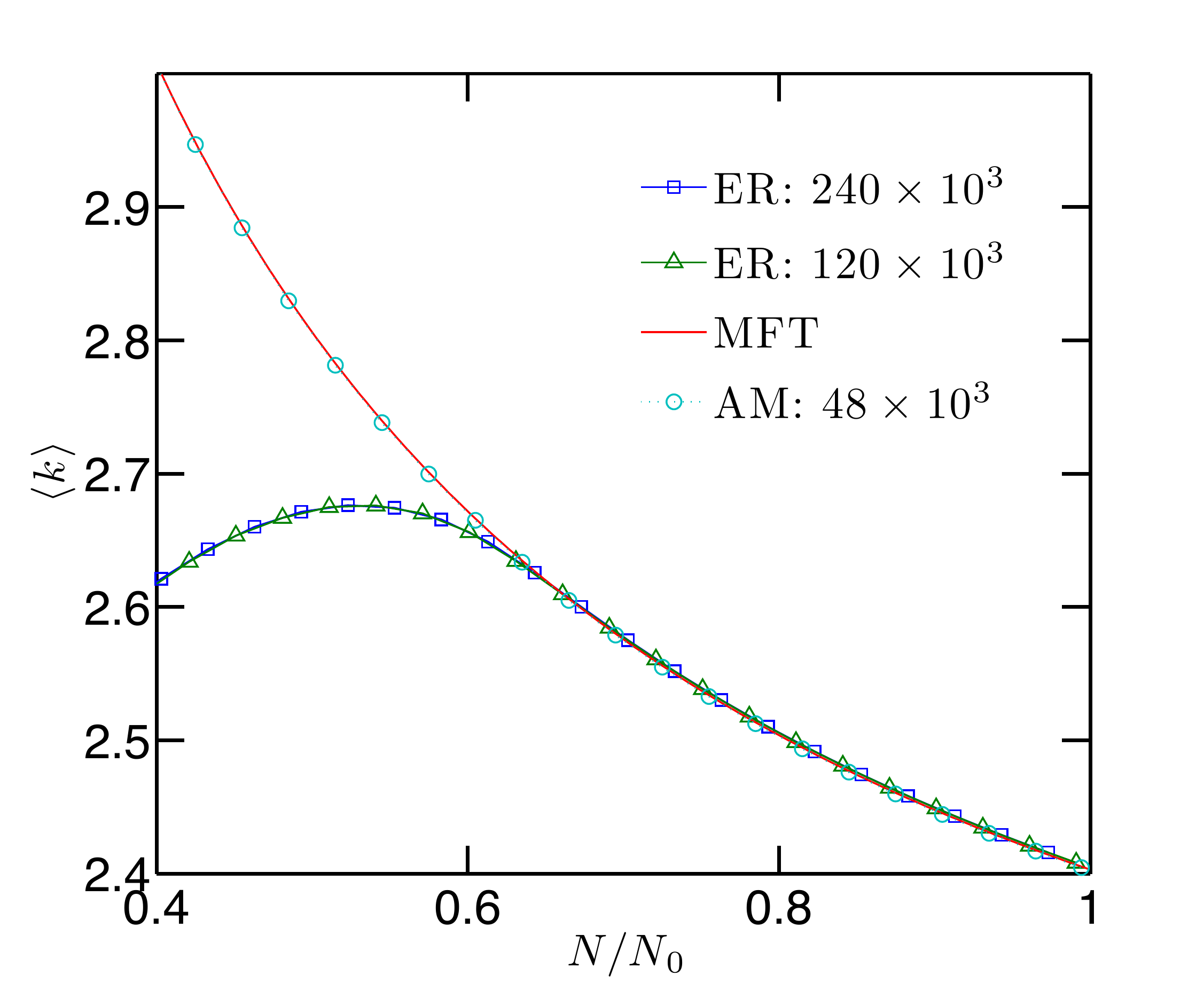}
\caption{(color online) Comparison between the annealed model (AM), mean-field 
   theory (MFT), and ER graphs with $\langle k \rangle^* =2$. There is good 
   agreement between theory and data in the mean-field regime $x \geq x_c$. 
   After the transition the effect of loops in ER graphs can no longer be 
   ignored and results in smaller $\langle k \rangle$ for ER graphs. }
\label{fig:k}
\end{figure}

The moments of the distribution can be obtained from 
\be
 \langle k^m \rangle = \left [ \left (x \frac{d}{dx}\right )^m G(x)\right ]_{x=1} \;.
\ee
One can check that the time derivative of the zeroth moment is zero, i.e. 
normalization is correct. The time derivative of the first moment is given by
\be
{d\langle k \rangle \over dt} ={1\over N} \left [\langle k\rangle ^2 -2 \langle k \rangle \right ] \;.
\label{eq:dkdt}
\ee
Using Eq.~(\ref{eq:Nt}) to convert to the derivative with respect to $N$ (for subtleties 
in this see \cite{tree_paper}) and integrating gives
\be
\langle k \rangle =  {a N_0 \over N} + 2 \;,
\label{eq:kt}
\ee
where $a=\langle k \rangle _0 -2$ and the subscript zero refers to the initial value.

To test the results of MFT we have simulated RSR for an {\it annealed random graph model} (AM) in 
the following way: We start with the degree sequence of the giant component of the ER graphs 
studied in the previous section, remembering for each of the $N_0$ nodes its degree, but remove 
all links. During each RSR step we first pick a random target node and read its degree $k$.
Then we pick $k$ other random nodes $i = 1,\ldots k$, this time with probabilities proportional 
to their degrees $k_i$. Finally we update the degree of the target to $k'=\left(\sum_1^k k_i\right)-k$ 
and discard the other $k$ nodes.

Figure~\ref{fig:k} compares Eq.~(\ref{eq:kt}) to the simulation results of the AM and of the model 
discussed in the last section starting with ER graphs. In all three cases we used $\langle k\rangle^*=2$. 
Due to loops in the ER graphs, the average degree of the ER graphs is always less than 
or equal to that of the AM or MFT. Note that ER graphs are locally tree-like and the effect of 
loops can be ignored initially. Thus before the transition -- in the mean-field regime -- there is 
complete agreement between the results of MFT, the AM, and the ER graphs. But after the 
transition, the effect of loops as well as fluctuations (which we will discuss later), result in a 
breakdown of the mean-field assumptions and the average degree of the ER graphs no longer 
agrees with the other two cases. 
 
In the mean-field regime the system size, $N$, can be found as an implicit function of $t$ by 
using Eqs.~(\ref{eq:Nt}) and~(\ref{eq:kt}) to get
\be
    t = -\frac{1}{2} \left \{ N-N_0 - {a N_0 \over 2} \ln\left[{a+2N/N_0\over a+2}\right]\right\} \;\;.
\ee
This result is shown in Fig.~\ref{fig:Nt} and is in good agreement with simulation results in 
the mean-field regime.

\subsection{Divergence of degree fluctuations}

\begin{figure}
\includegraphics[width=1\columnwidth]{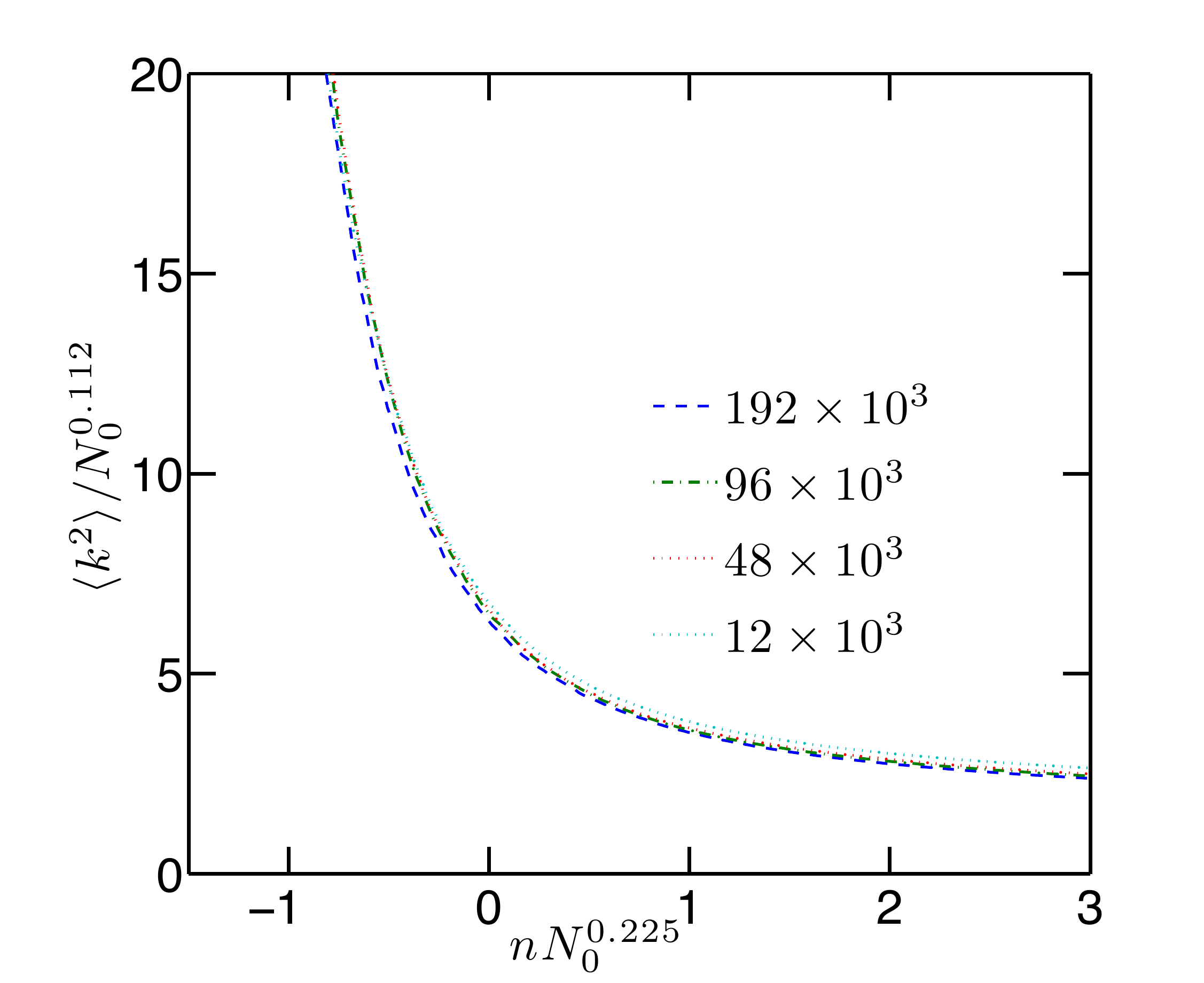}
\caption{(color online) FSS analysis of the second moment of the degree distribution, 
   $\langle k^2\rangle$, close to criticality for the AM. The values of $x_c=0.718$, and 
   $\gamma=0.5$ used in the plot are those obtained from the MFT. 
   While the exponent $1/\nu=0.225\pm 0.015$,  obtained from the FSS data collapse is 
   similar to ER graphs, the exponent $\gamma$ is different from $\gamma=0.88$ in ER graphs. }
\label{fig:k2_mean_scale}
\end{figure}

Also within MFT, the variance of the degree distribution diverges at the transition point.
For ease of calculations we switch to factorial moments of the degree distribution obtained by 
consecutive derivatives of the generating function,
\bea
    &\langle k_-^m \rangle &=\langle k(k-1)...(k-m+1) \rangle \nonumber\\ 
    &&=\left [ \left ({d \over dx}\right )^m G(x)\right ]_{x=1}  \;\; .
\label{eq:fact}
\eea

Using Eq.~(\ref{eq:dGdt}), the time derivative of the second factorial moment is
\bea 
&\displaystyle {d\over dt}\langle k_-^2\rangle =&{\langle k_-^2\rangle \over N}
   \left [ {\langle k_-^2\rangle^2\over \langle k \rangle^2} 
   +\langle k \rangle -3  \right] \;\;.
\label{eq:dk2dt}
\eea
We next define a variable $u=\langle k_-^2\rangle/ \langle k \rangle$,
and use Eqs.~(\ref{eq:Nt}),~(\ref{eq:dkdt}) and (\ref{eq:kt}) to get
\be
{du \over u^3-u}= -{dN \over aN_0+2N} = -{dx \over a+2x}
\ee
with $x=N/N_0$. Integrating this equation leads to 
\be
u^2=\frac{a+2x}{a-c+2x} \;\;,
\label{eq:u2}
\ee
where
\be
c=\frac{u_0^2-1}{u_0^2}(a+2) \;\;,
\label{eq:c}
\ee
and $u_0$ is the initial value of $u$. Since the average degree $\langle k \rangle$ does 
not diverge at the transition, the divergence of $u$ is the same as the divergence of the 
variance of the degree distribution. The quantity $u$ diverges when the denominator of 
Eq.~(\ref{eq:u2}) vanishes, so the critical point is at
\be
x_c={N_c\over N_0}={1\over 2}\left[c-(\langle k\rangle_0-2)\right] \;\;.
\label{eq:xc}
\ee
Equations~(\ref{eq:c}), and~(\ref{eq:xc}) result in $x_c=0.718 \ldots$ for the AM model 
with $\langle k\rangle_0 = 2.4$ we study here (notice that the initial degree distribution
is not strictly Poissonian due to the restriction to the giant component of the original
ER graph). 
Substituting  $n= (x-x_c)/x_c$ into Eq.~(\ref{eq:u2}) we get

\be 
\langle k^2\rangle  \sim u \sim n^{-\gamma} \quad    {\rm with }\quad \gamma=1/2 \;\;.
\label{eq:k2scale}
\ee
For a finite system, we make the FSS ansatz
\be
{\langle k^2\rangle \over N_0^{\gamma/\nu}}=f(nN_0^{1/\nu})\;\;.
\ee
Figure~\ref{fig:k2_mean_scale} shows an FSS analysis of $\langle k^2\rangle $ for 
the annealed model, close to criticality, and for several system sizes. The values 
of $x_c=0.718$ and $\gamma=0.112/0.225=0.5$ used in the FSS analysis are taken 
from MFT and give excellent agreement.
\\

\subsection{Other critical exponents}

To get all other exponents in MFT we use Eqs.~(\ref{eq:dGdt}) and~(\ref{eq:fact}) to 
find the time derivative of the third factorial moment, $h=\langle k_-^3\rangle$
\be
   \frac{dh }{dt}=\frac{hu^3}{N} + \ldots\;\;.
\ee
where the dots stand for terms that are less divergent for $x\to x_c$. Together
with $du/dt \sim u^3/N +\ldots$ this gives near the critical point
\be
   \ln h\sim u\sim n^{-1/2} \Rightarrow h\sim e^{1/n^{1/2}}\;\;,
\label{eq:h}
\ee
suggesting that the third moment has an essential singularity.
The latter seems to be contradictory to scaling theory, but it really is not, and 
there exists a consistent solution showing these features. Assume the scaling ansatz
\be
   p_k=k^{-\tau} f(k/k_{\rm cutoff}) 
\ee
for the degree distribution near the critical point, with $k_{\rm cutoff}$ diverging
at $x\to x_c$. For $u$ to diverge, $\tau$ must
be $\leq 3$. If $\tau$ were strictly $<3$, we would have $u\sim k_{\rm cutoff}^{3-\tau}$
and $h\sim k_{\rm cutoff}^{4-\tau}$, i.e. there would be a power relation between 
them: $h\sim u^{(4-\tau)/(3-\tau)}$. The only way to obtain $u\sim \ln h$ is by 
having a logarithmic divergence of the sum $\sum_k k^2p_k$, i.e. 
\be
   \tau=3\;\;.
\ee 
In order to have $\gamma=1/2$, one needs furthermore $k_{\rm cutoff}\sim e^{1/n^{1/2}}\sim N_0$, 
giving then also Eq.~(\ref{eq:h}). The fact that $k_{\rm cutoff}$ diverges faster
than a power for $x\to x_c$ means that there is no simple scaling theory near the transition due to the singularity.



\subsection{Limiting behavior for ${\bf\langle k\rangle ^*\to 1}$ }

In the limit $\langle k \rangle^*=1$, the giant components of ER networks become trees 
with $\langle k\rangle _0=2$. Since trees remain trees during RSR, $\langle k \rangle =2$
during the entire RSR flow, consistent with Eq.~(\ref{eq:dkdt}). On the other hand, 
$\langle k^2 \rangle$ does increase with $t$. Equation~(\ref{eq:u2}) leads to
\be
   u^2={2x\over 2x-c}\;\; ,
\ee
and Eq.~(\ref{eq:xc}) gives
\be
   x_c={u_0^2-1\over u_0^2}>0\;\;.
\ee

This is in contrast to the result of \cite{tree_paper}, where we found $x_c=0$ for critical 
trees. Indeed, the limit $\langle k \rangle^*\to 1$ of the present model is {\it not} 
the model of critical trees that was treated in \cite{tree_paper}.

This follows from how the critical trees of \cite{tree_paper} and critical ER graphs 
are generated. In ER graphs links are distributed among nodes completely at random. If a 
node is picked at random, the degrees of all its neighbors are 
distributed according to 
\be
   k p_k/\sum_l lp_l,    \label{neighbrs} 
\ee
and there is no further structure. In contrast, the critical trees of \cite{tree_paper} are 
generated by a critical random branching process that starts from one particular node and 
imprints on them a {\it rooted} structure. Therefore, if a node is picked randomly, there are
relations that hold seperately for its mother and its daughters. While the degree distribution 
for the mother satisfies Eq.~(\ref{neighbrs}) with $k$ replaced by $k-1$, the degree 
distribution of the daughters is simply $p_k$. One might think that this subtle difference can 
be neglected in a mean-field approximation, but this is not true: Since each RSR step affects
three generations of nodes, a consistent grandmother-mother-daughter relationship has an 
effect on the RSR flow. But it is not intuitively clear why this small difference has such 
a strong influence on the threshold for AP. Notice that an even more surprising dependence
on minor details, leading indeed to a violation of universality, is seen also in AP on 
2-dimensional lattices~\cite{claireAP}.


We did not study the case $\langle k \rangle^*=1$ numerically, because the size of the 
largest component in critical ER graphs of size $N_0$ is $\sim N_0^{2/3}$, making it very 
difficult to create large initial connected graphs.

\section{Fluctuations in the hub phase}

If the giant cluster (or hub) is itself a target of RSR, the size of the network decreases 
significantly in that time step. This gives rise to large fluctuations in the size of the network. 

\begin{figure}
\includegraphics[width=1\columnwidth]{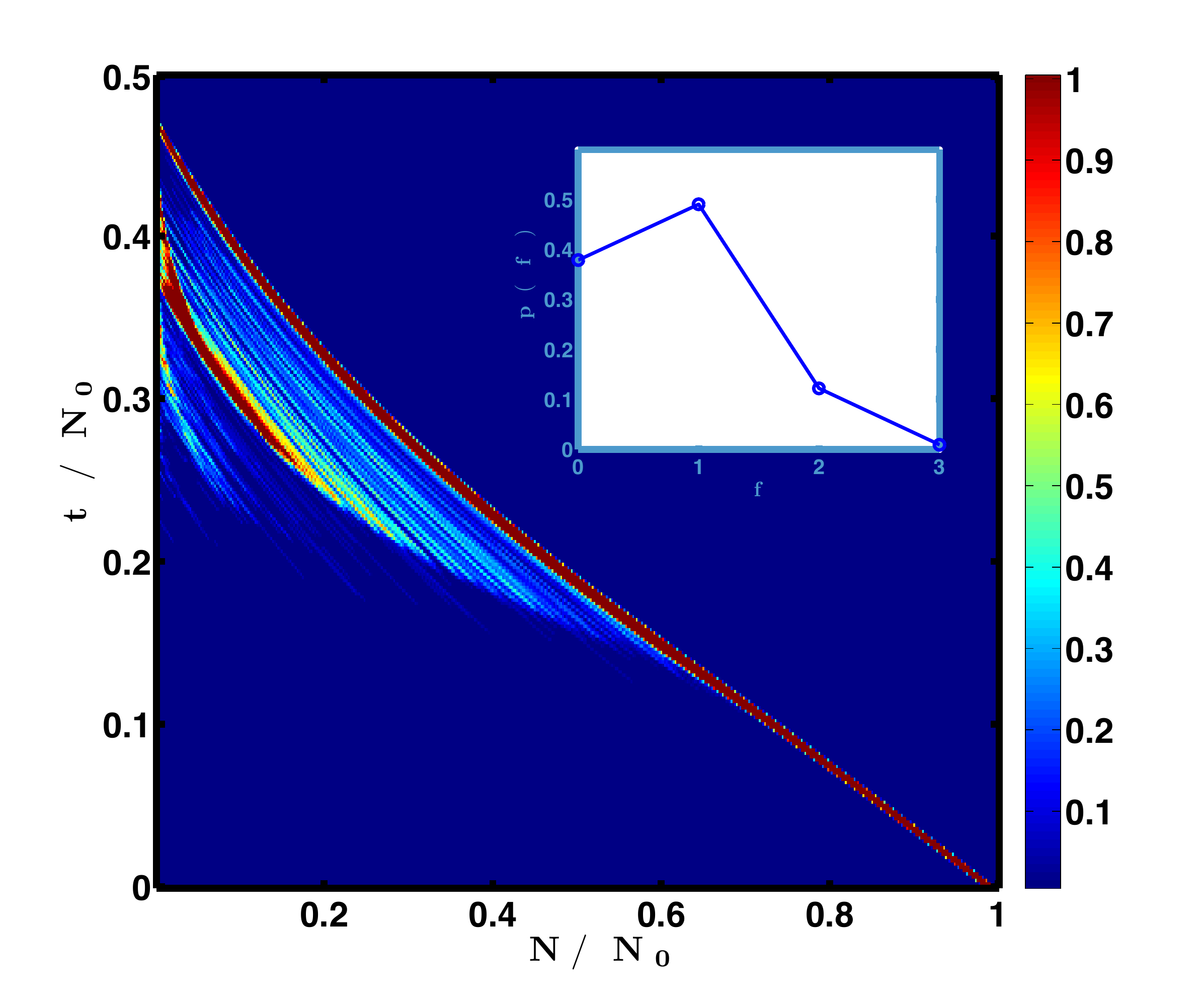}
\caption{(color online) Scatter plot of $t$ {\it vs.} $N$ in rescaled units for ER graphs 
   with $\langle k \rangle ^*=2$ and $N^*=120\times 10^3$. The color map shows the relative frequency 
   of each $(N,t)$ pair in the ensemble. The main, intermediate and weak bands correspond 
   to realizations where the giant hub has been hit zero, one or two times, respectively. 
   The inset shows the probability that the giant hub is targeted $f$ times by RSR. }
\label{fig:scatter}
\end{figure}

Figure~\ref{fig:scatter} shows a scatter plot of $t$ {\it vs.} $N$ in rescaled units, for an 
ensemble of ER networks with $N^*=1.2\times 10^5$ and $\langle k\rangle^*=2$. The $x$ and $y$ axes in this plot are 
coarse grained into $500$ and $200$ bins respectively, giving $100,000$ pixels. The color of each pixel represents the 
frequency of this $(N,t)$ pair relative to the frequency of the most populated pixel. 
As one can see, an envelope exists corresponding to the largest $N$ at a given time (and biggest 
time for a given N). There is also a second band (of high probability) which corresponds 
to intermediate $N$. A third but weak band also appears at smaller system sizes, which 
is more difficult to distinguish due to the considerable fluctuations.

The inset of Fig.~\ref{fig:scatter} shows the fraction of realizations in which the giant hub 
is hit $f$ times, conditioned on $k_{\rm max}/N_0>0.1$.  Most often the hub is hit only once, and never more than three times. Evidently, the envelope (the uppermost band)
comes from the realizations in which the giant hub was not hit at all, the intermediate band
results from cases where the giant hub is hit once, and the third weak band is due to rare 
cases where the giant hub is hit twice.

The slopes of the main bands are also informative. The uppermost band starts with slope 
$\frac{-1}{\langle k \rangle_0}$, in agreement with Eq.~(\ref{eq:Nt}).
At final stages, where the structure is star-like (as discussed in the following section), the 
bands have slope $-1$, which means that in most cases a leaf is targeted and thus one node is 
removed in one time step. The wide range of values for realizations as shown in 
Fig.~\ref{fig:scatter} explains why in Fig.~\ref{fig:Nt} averaging over $t$ at fixed 
$N$ gave a different result than averaging over $N$ at fixed $t$.

The distribution of times $T$ for the networks to reach $N=1$ corresponds to the leftmost column 
in Fig.~\ref{fig:scatter}. This distribution has a shoulder where the uppermost band hits the y-axis, 
(at $t/N_0\approx 0.47$) and a peak where the second one hits it ($t/N_0\approx 0.38$). These
distributions show perfect data collapses for different system sizes (data not shown). For networks 
with $\langle k\rangle ^*>2$ the shoulder turns into a second peak which grows and becomes the 
dominant peak on increasing $\langle k \rangle^*$. It should disappear for $ \langle k\rangle ^*\to 1$.

\subsection{Scaling behavior at late times}

\begin{figure}
\includegraphics[width=1\columnwidth]{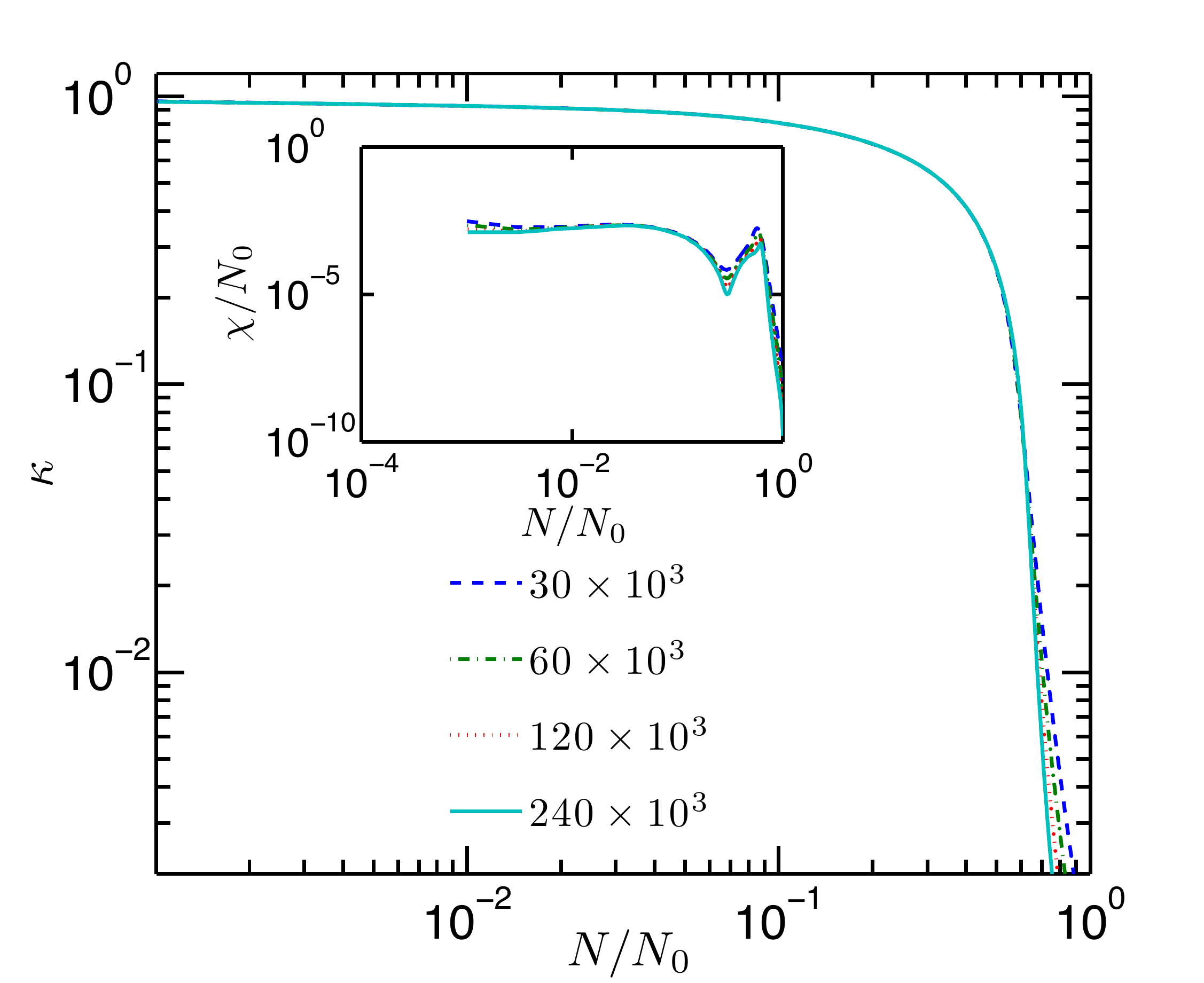}
\caption{(color online) The relative maximum degree, $\kappa=k_{\rm max}/(N-1)$, {\it vs.} $N/N_0$
   for different system sizes. The plot shows that the network is star-like at late times since 
   $\kappa$ approaches $1$. Inset: Variance of $\kappa$ {\it vs.} $N/N_0$. Although the qualitative 
   behavior of graphs under RSR at late times is the same as for the (quasi-)parallel renormalization 
   method~\cite{Radi1}, the quoted exponents are different.}
\label{fig:kappa}
\end{figure}

Eventually as the networks shrinks, $k_{\rm max}$ starts to decrease and at the same time the 
network topology moves towards a star-like structure. 

The relative size of the largest hub $\kappa=k_{\rm max}/(N-1)$ is a good measure for the 
similarity of a graph to a star which is a graph whose nodes are at most a distance two apart.  
Figure~\ref{fig:kappa} shows $\kappa$ and its variance $\chi$ as a function of the relative 
system size. As one can see, at late stages of RSR $\kappa$ is close to one, and thus the network 
has a star-like structure.   

The star-like regime was also observed in previous (quasi-)parallel methods used for 
renormalizing networks~\cite{Radi1, Radi2}. Comparing our analysis with those studies, 
RSR shows scaling and criticality in the flow at early times that was not picked up 
previously, because the renormalization steps in the quasi-parallel method were too large 
and jumped over the agglomerative percolation transition. Thus only the scaling at late times was observed in 
\cite{Radi1, Radi2}. Although the graphs under RSR look qualitatively similar to those obtained
with the quasi-parallel method at late times, the quoted exponents are different 
(our Fig.~\ref{fig:kappa} should be compared with Fig.~1 in Ref.~\cite{Radi1}).

\subsection{The star regime}

\begin{figure}
\includegraphics[width=1\columnwidth]{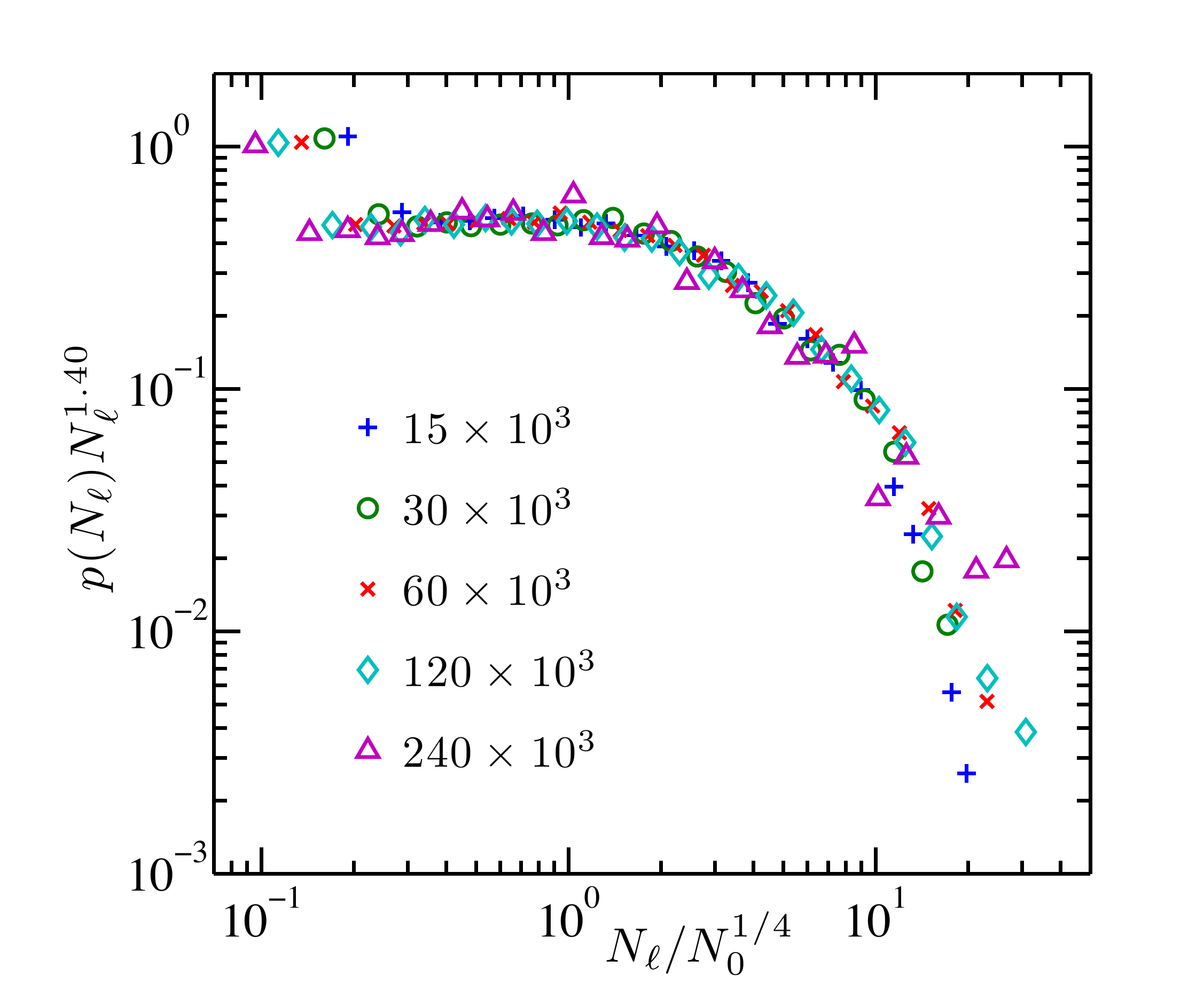}
\caption{(color online) Data collapse for the distribution of the last sizes. The distribution 
   follows the FSS ansatz in Eq.~(\ref{eq:star}) with $\tau_s=1.40\pm 0.15$ and $D_s=0.25\pm 0.05$,
   except for the leftmost points. The reason for their special behavior is given in 
   \cite{tree_paper}.}
\label{fig:Nlast}
\end{figure}

We define $N_{\ell}$ to be the last size of the network one step before it collapses into a single node.
By definition the network has to be a pure star at this point. Figure~\ref{fig:Nlast} 
shows a data collapse for the distribution of $N_{\ell}$ for ER graphs of different 
sizes. It is a broad distribution following the scaling ansatz
\be
p(N_{\ell}) \sim \frac{1}{N_{\ell}^{\tau_s}} f(\frac{N}{N_0^{D_s}}) \quad,
\label{eq:star}
\ee
with $\tau_s=1.40\pm 0.15$ and $D_s=0.25\pm 0.05$.

The exponents $\tau_s$ and $D_s$ are similar to the ones obtained for critical trees~\cite{tree_paper}. This suggests universality in the final structure of the graphs, regardless of the starting structure, as the graph collapses into a single node and all original structure is lost.

\section{starting with other average degrees}

\begin{figure}
\includegraphics[width=1\columnwidth]{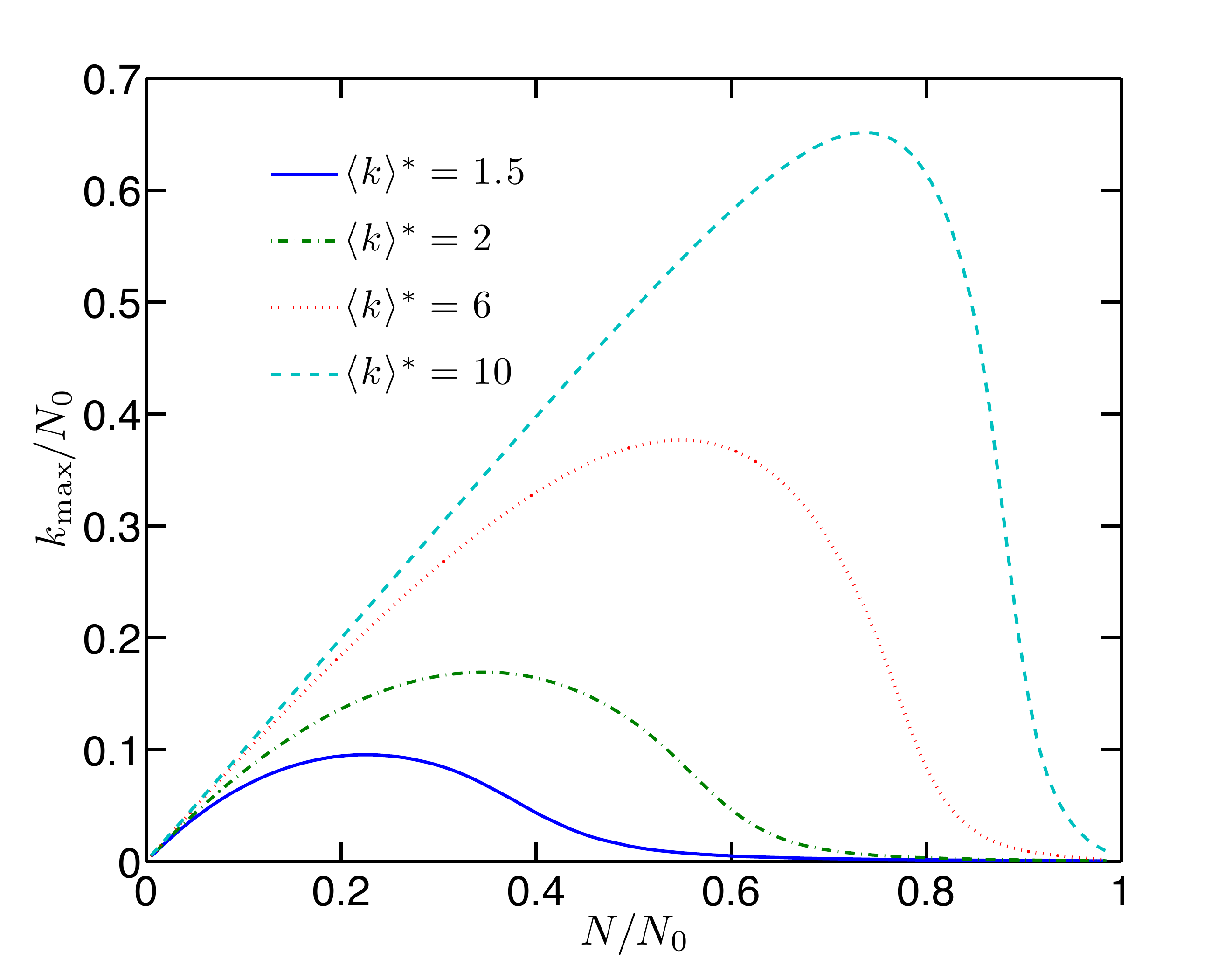}
\caption{(color online) Rescaled maximum degree, $k_{\rm max}/N_0$ {\it vs.} system size
    for ER graphs with $N^*=30000$. The transition shifts to the right with increase of 
    $\langle k \rangle^*$.}
\label{fig:kmax-k0}
\end{figure}

Up to now we studied the behavior of ER graphs with $\langle k\rangle^*=2$. Here we 
discuss the effect of the initial average degree on RSR flow, still considering ER
graphs. 

Figure~\ref{fig:kmax-k0} represents $k_{\rm max}$ for ER graphs with different values 
of $\langle k \rangle^*$.  The figure demonstrates similar critical behavior for these networks.
For higher initial average degree the transition gets sharper and moves to the right, i.e. 
towards earlier times. This is also predicted from Eq.~(\ref{eq:xc}). 
For larger $\langle k \rangle^*$, $x_c$ approaches $1$. Note that both $\langle k\rangle_0$ 
and $\langle k^2\rangle_0$ affect the position of $x_c$.

\begin{figure}
\includegraphics[width=1\columnwidth]{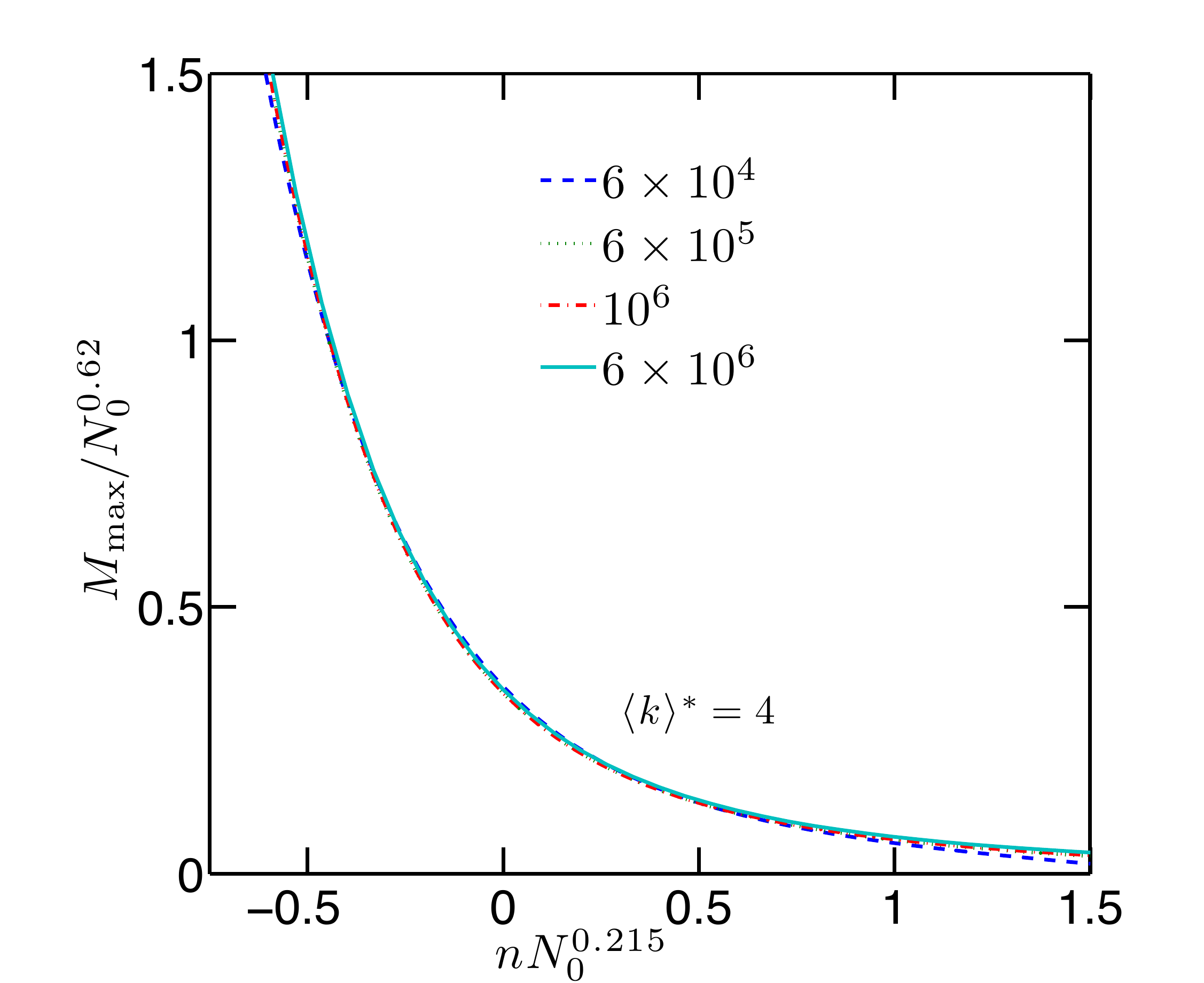}
\caption{(color online) Scaling of $M_{\rm max}$ for ER graphs with $\langle k \rangle^*=4$. 
   The values $x_c=0.865\pm 0.010$, $1/\nu=0.215\pm0.030$ and $1-\beta/\nu=0.62\pm0.05$, 
   obtained by finding the best data collapse agree with Eq.~(\ref{eq:xc}) 
   and the exponents for ER graphs with $\langle k \rangle^*=2$ within error bars. }
\label{fig:mscale_k4}
\end{figure}

Figure~\ref{fig:mscale_k4} shows an FSS analysis of $M_{\rm max}$ for ER graphs with 
$\langle k \rangle^*=4$. The critical point $x_c=0.865\pm 0.010$ and the exponents 
$1/\nu=0.215\pm 0.030$ and $1-\beta/\nu=0.62\pm 0.05$ are obtained by finding the best 
data collapse. The value of $x_c$ is in agreement with Eq.~(\ref{eq:xc}) and 
the exponent $\nu$ and $\beta$ agree with those for $\langle k\rangle^*=2$, 
within our error estimates. For even higher average degrees (not shown) the exponents still 
agree with the ones obtained for $\langle k\rangle^*=2$, although the error bars are rather 
large.

\section{RSR with larger box sizes}

\begin{figure}
\includegraphics[width=1\columnwidth]{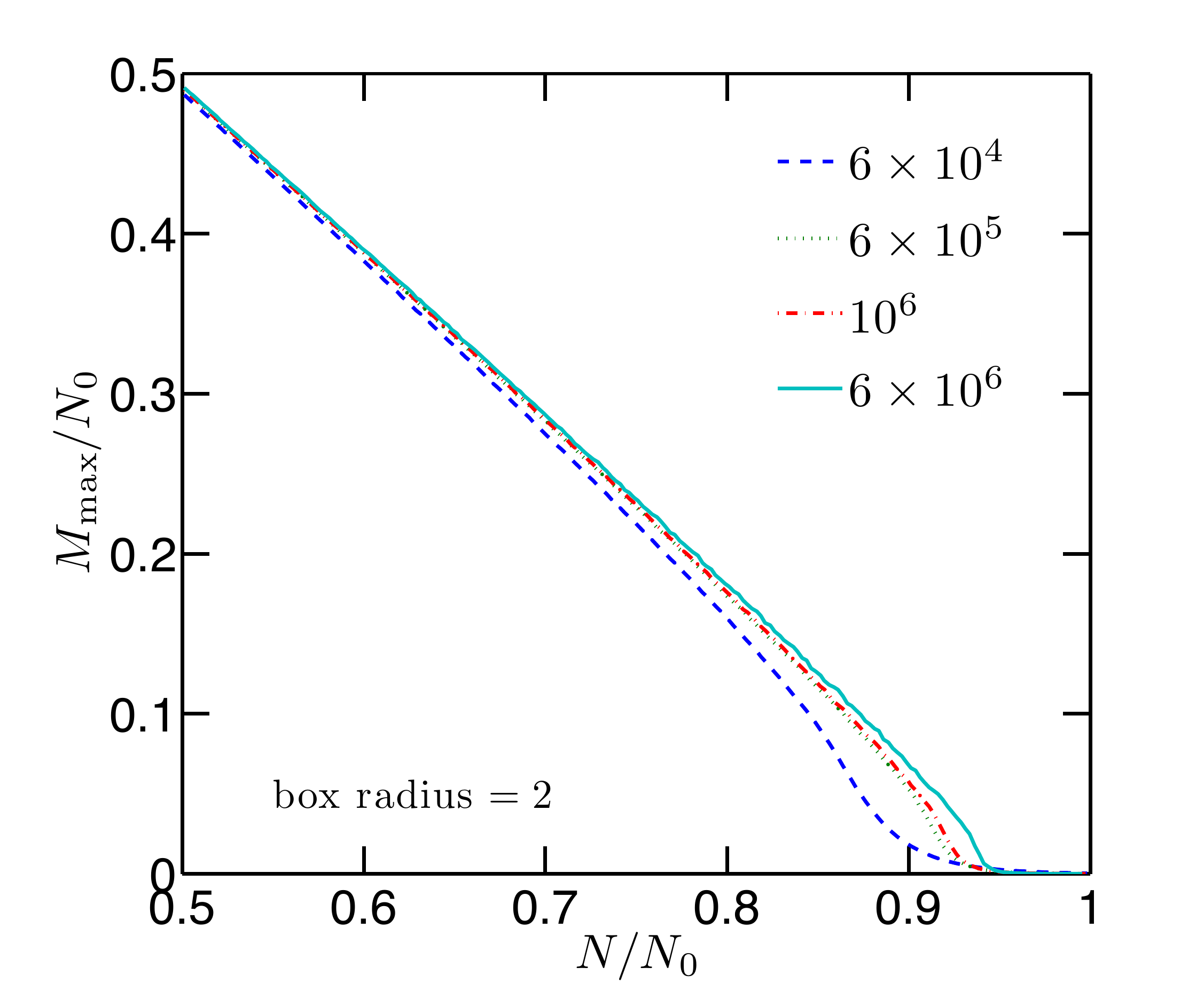}
\caption{(color online) Flow of the order parameter under RSR with box radius $b=2$ for 
   ER graphs of $\langle k \rangle^*=2$ and several sizes. The data shows a sharp 
   transition at early times, but a clean FSS analysis including precisely locating the critical 
   point is not numerically tractable. }
\label{fig:mmax-box}
\end{figure}

In this section we study RSR with box radius $b>1$ on ER graphs with $\langle k \rangle^*=2$.
We see evidence for a transition at early times.
Figure~\ref{fig:mmax-box} shows the order parameter as a function of $N/N_0$ for 
networks of different system sizes under RSR with $b=2$. Although one can clearly see 
evidence for a phase transition at early times, extrapolating the critical point for 
the infinite system with precision is not possible.  

One of the main differences between the $b=1$ case and $b>1$ is that for larger box radii there 
is no star regime. Once the network has diameter two, it will die in the next step with 
probability one. 

Another point to mention is that with any box size larger than one, the possibility to 
incorporate the hub at any step is large. The reason is that RSR with $b>1$ is performed 
by targeting the same node $b$ successive times.  Although the target itself is not likely 
to be the hub, it is likely that it is the neighbor of the hub and thus merges with it. 
Hitting the same node again means then hitting the hub with high probability. With this 
argument any box size higher than one is similar to a {\it weighted} RSR, where nodes are 
being targeted with probability proportional to their mass or degree.  

\section{Scale-free networks}

\begin{figure}
\includegraphics[width=1\columnwidth]{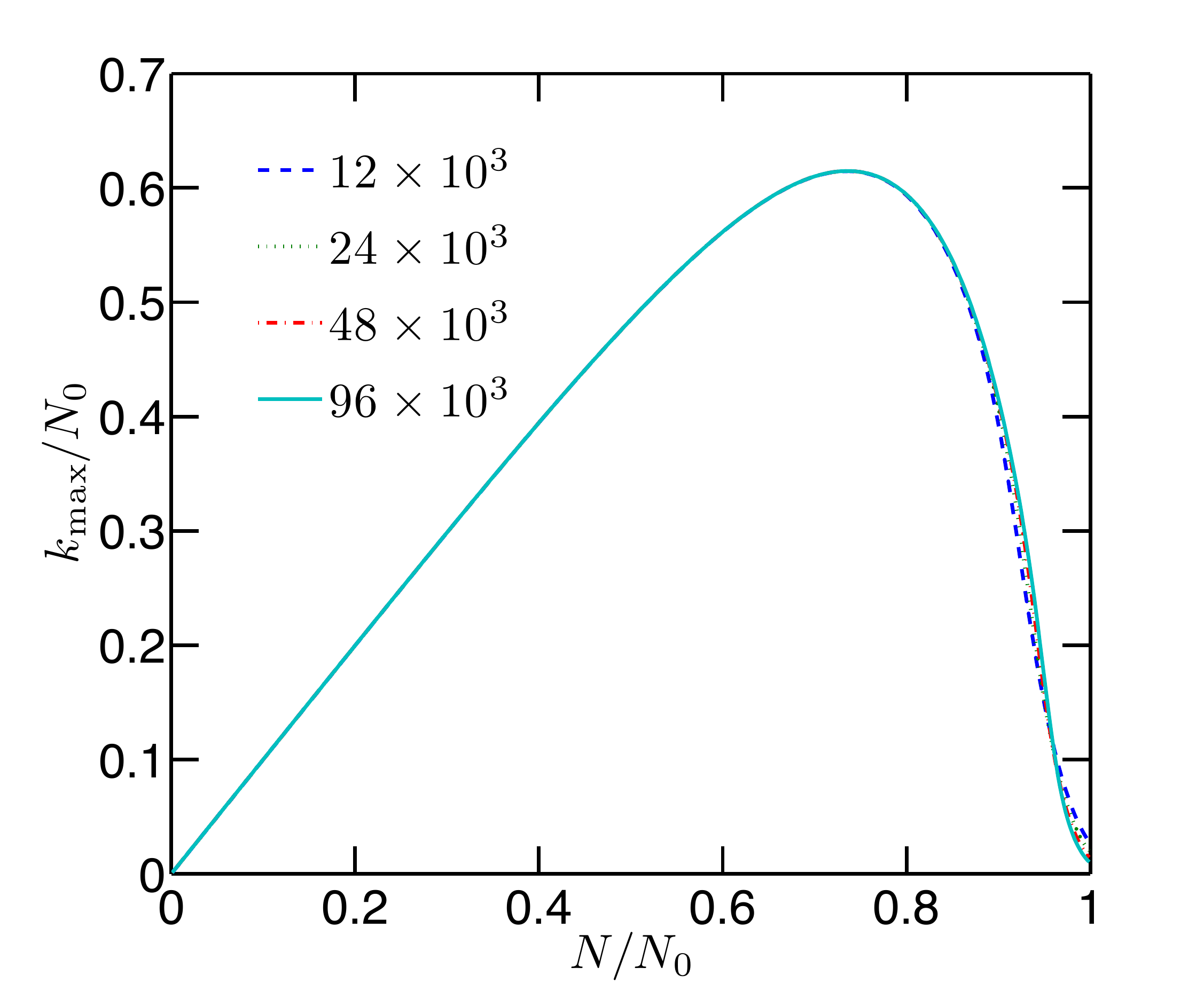}
\caption{(color online) FSS analysis of $k_{\rm max}/N_0$ for the BA model. The critical point 
   is pushed towards one, and there is a perfect data collapse after the hubs are well established.}
\label{fig:BAkmax}
\end{figure}

\begin{figure}
\includegraphics[width=1\columnwidth]{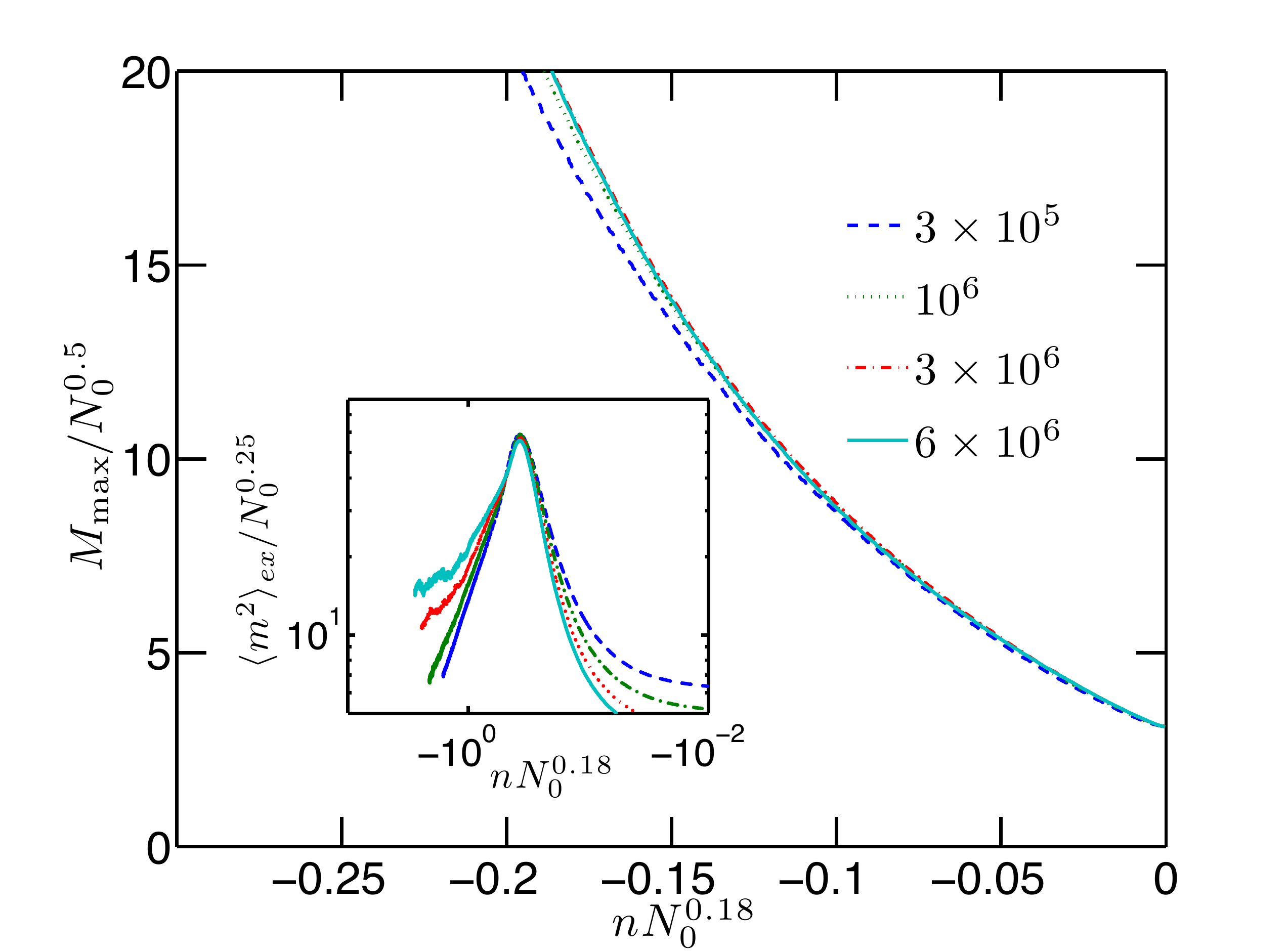}
\caption{(color online) FSS analysis of $M_{\rm max}$ and $\langle m^2\rangle_{ex}$ for the BA 
   model. The critical point is set at $x_c=1$. The exponents $1/\nu=0.18\pm0.02$, $D=0.5\pm0.1$ 
   and $\gamma/\nu=0.25\pm0.03$ are obtained from the data collapse.}
\label{fig:BAMmax_scale}
\end{figure}

Models with broad or ``scale-free'' degree distributions are often more interesting in view of their 
application to real-world networks. We have studied  RSR on the 
Barabasi-Albert model~\cite{BA}. Figure~\ref{fig:BAkmax} shows the behavior of the 
maximum degree under RSR. Since these networks are scale free at the beginning, 
the transition is pushed all the way to $N/N_0=1$. There is perfect data collapse 
after the hubs are well established.

The critical point can also be obtained from Eq.~(\ref{eq:xc}). The value of $\langle k^2\rangle_0$ --
and thus also the value of $u_0=\langle k_-^2\rangle_0 / \langle k \rangle_0$ -- diverge for 
scale-free networks, giving $x_c\approx 1-1/u_0^2 \approx 1$.

When analyzing mass distributions for renormalized scale-free networks, it can be argued 
that one should not give masses $m=1$ to all nodes of the initial graph. Instead one might
assign to every node a mass equal to its degree, as this allows one to consider mass as a 
proxy for the degree of nodes in the simulation of the RG flow. 

This convention is used in Figure~\ref{fig:BAMmax_scale}, which illustrates an FSS analysis for
the maximum cluster mass, $M_{\rm max}$, and the second moment of the mass distribution
excluding the largest cluster, $\langle m^2\rangle_{ex}$, in BA networks of several sizes. 
Setting the critical point at $x_c=1$, we obtained $1/\nu=0.18\pm0.02$, $D=0.5\pm0.1$ and $\gamma/\nu=0.25\pm0.03$.

\section{conclusion}

In this paper we have extended Random Sequential Renormalization (RSR) to 
several networks, namely Erd\"os-R\'enyi and scale-free networks. In each step of RSR 
only a local part of the network within a fixed distance from a randomly chosen node is 
coarse-grained into one node. This is in contrast to (quasi-)parallel RG schemes that tile and 
coarse-grain the whole network in one step -- which, however, has to be broken
up into sequential local sub-steps for technical reasons.  
Apart from simplicity of the algorithm, RSR generates considerably larger amount of statistics 
and allows for a more detailed analysis of the renormalization flow. RSR can be interpreted as a 
cluster growth process where at each step a randomly chosen cluster grows at its boundary by 
agglomerating to all its neighboring clusters. Hence, the fast Monte Carlo algorithm of Newman and 
Ziff~\cite{newman_ziff1,newman_ziff2} for percolation can be used to simulate RSR on networks of 
up to millions -- or even billions -- of nodes. 

For all the graphs we studied, RSR leads to a continuous agglomerative percolation transition (AP) where the 
largest cluster (node) outgrows all others both in terms of its mass and degree. We found three 
universality classes (critical trees, sparse ER graphs and mean-field AP) for evolution of 
networks under RSR. For sparse ER graphs we derived the corresponding critical exponents 
numerically and found that the exponents obtained by analysis of the masses of the clusters are 
not different from the ones obtained by analyzing the degrees of the nodes. Since mass analysis 
can be performed much faster with the help of the NZ algorithm, we suggest that mass analysis 
may be better suited to extracting scaling properties of large networks. Regardless of the initial 
average degree of the ER graph, we found the same critical exponents for the percolation 
transition, within error. At late stages of RSR, graphs experience a regime in which they 
switch to a star structure for $b=1$. For both ER graphs and critical trees this regime extends in the 
range $1<N<N_0^{1/4}$. 
  
For scale-free networks the transition is forced to $x_c=1$.  Hence our data collapse methods for 
finding the critical exponents of scale-free networks are not as neat as for ER graphs, and 
this makes it hard to decide whether BA and ER networks are in the 
same universality class. 

While the scaling behavior of critical trees under RSR is similar to graph behavior under 
the (quasi-)parallel renormalization scheme studied by Radicchi {\it et. al.}~\cite{Radi1, Radi2}, 
the percolation transition revealed by our method in the early stages of the RG flow 
is not {\it seen} in their analysis. We conjecture that it exists also there in principle,
but it would be very hard to study due to the coarseness of their RG flow observation. At 
final stages RSR and parallel schemes lead to the same qualitative picture, namely a 
star-like structure for $b=1$, but the scaling behavior and the corresponding exponents are different. 

The simplicity of RSR as well as the fact that it is a percolation process both for networks 
and lattices makes it a useful tool for studying complex networks. For real-world networks 
finite-size scaling analysis is not generally possible since every network has a fixed (finite) size. 
But even in that case high statistics of RSR flow and the efficiency of the algorithm make 
it possible to study the scaling properties of individual large networks. 

\acknowledgements
We thank Claire Christensen and Seung-Woo Son for numerous discussions and helpful comments.

\appendix*
\section{relationship between mass and degree of nodes}

\begin{figure}
\includegraphics[width=1\columnwidth]{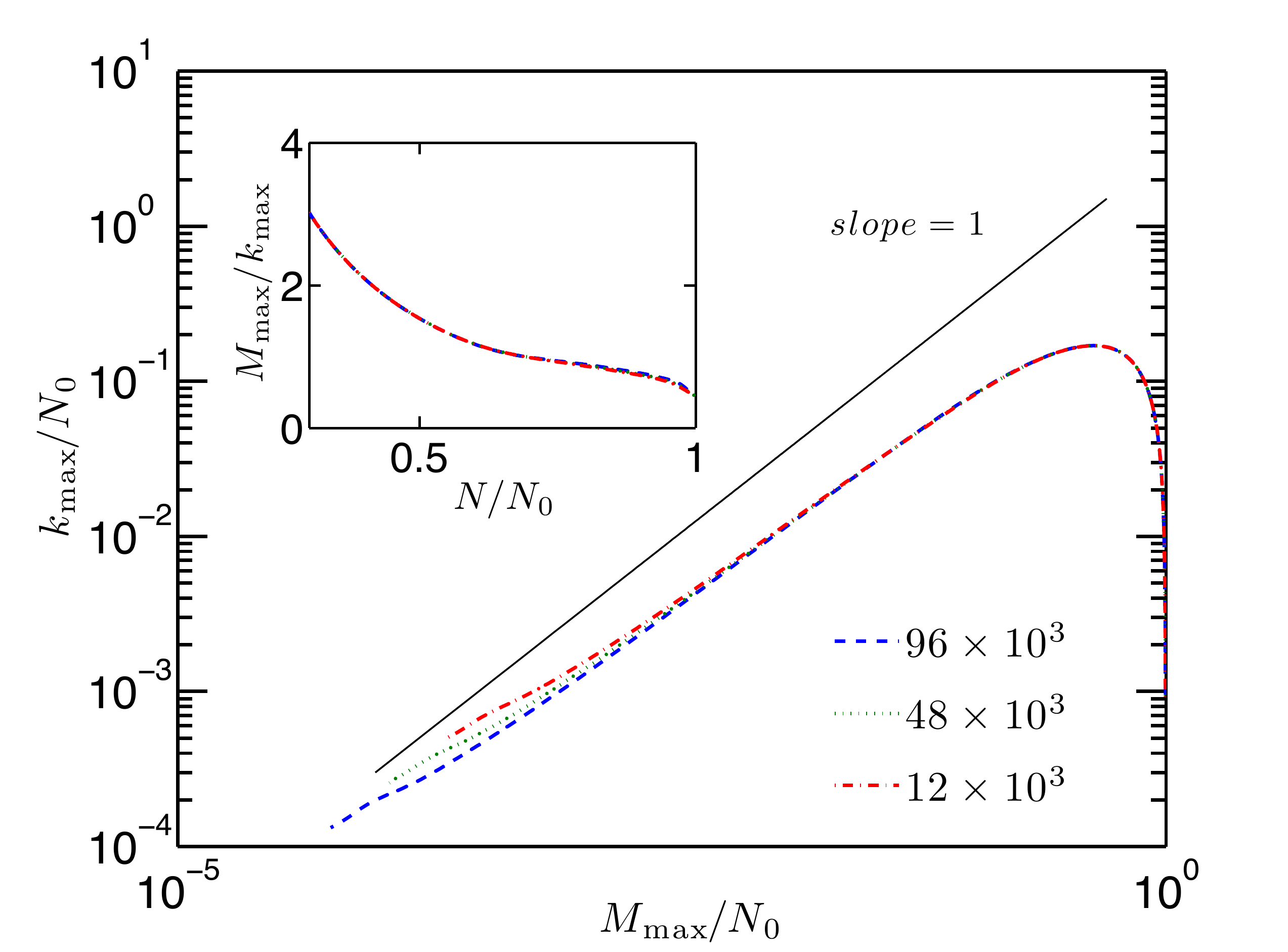}
\caption{(color online) Log-log plot of normalized maximum degree $k_{\rm max}/N_0$ {\it vs.} 
   $M_{\rm max}/N_0$ for ER graphs with $\langle k \rangle^*=2$ and size $N^*=1.2\times 10^5$. These are proportional to each other in a region close to criticality which 
   extends over larger domains with increase of system size. While $M_{\rm max}$ increases with 
   $N_0$ monotonically, $k_{\rm max}$ is confined to the current system size $N$ and starts to decrease 
   deep in the hub phase. The inset shows the ratio of maximal mass and degree on a linear 
   scale. The curve is linear in the critical region. }
\label{fig:km_Mm}
\end{figure}

Figure~\ref{fig:km_Mm} depicts the linear relation between $k_{\rm max}$ and $M_{\rm max}$ in 
the critical region for ER graphs with $\langle k\rangle^*=2$ and several system sizes. This 
shows that either of them can be used to extract renormalization flow properties near the transition. Since RSR can be simulated much faster if we only measure the mass-related quantities 
(instead of degree), we suggest that the RG analysis in the critical region can be confined 
to mass-related quantities. In the final stages of the flow ($N/N_0\lesssim 0.3$), $k_{\rm max}$ 
decreases as it cannot exceed the number of nodes present in the system. $M_{\rm max}$, on the 
other hand, increases monotonically till the end of the process where $M_{\rm max}=N_0$. 

\begin{figure}
\includegraphics[width=1\columnwidth]{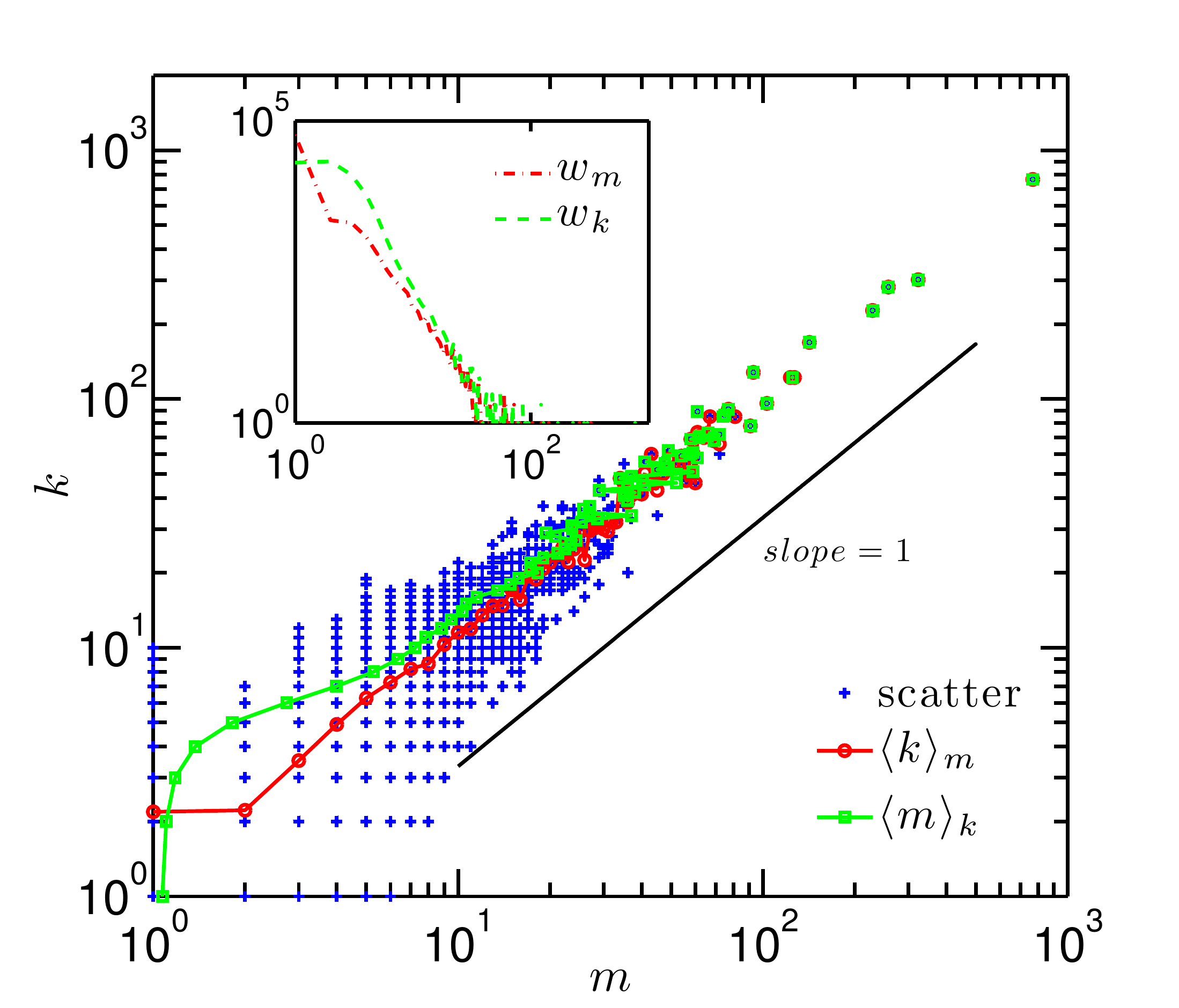}
\caption{(color online) Scatter plot of  masses and degrees of all clusters close to 
   criticality. The data is obtained from one RSR trajectory of an ER graph of $N^*=1.2\times 10^5$  
   nodes, with $\langle k\rangle^*=2$ at $N/N_0=0.688$. The red line with circles shows 
   the average degree of clusters of a given mass, and the green line with squares shows the 
   average masses of nodes with a given degree. The inset shows the number of clusters of a 
   given mass ($w_m$) and the number of nodes of a given degree ($w_k$). }
\label{fig:Mk}
\end{figure}
The correlation between mass and degree of clusters close to criticality is shown in 
Fig.~\ref{fig:Mk} for ER graphs of $N^*=1.2\times 10^5$ at $N/N_0=0.688$. Each point in the 
scatter plot shows one $(m,k)$ pair in the whole network. Also shown are the average degree 
$\langle k \rangle_m$ of nodes of a given mass and the average mass $\langle m\rangle_k$ of 
clusters with a given degree. In the inset we show the number of clusters with a given mass, 
$w_m$, and the number of nodes with a given degree, $w_k$.

The average mass of nodes with degree one is close to 1, which means that most of them have 
not been hit by RSR, and the average degree of clusters with mass one is more than two which 
means that the nodes that have not been hit by RSR keep their starting average degree. 
Since the two averages differ only for masses (degrees) less that ten, we suggest that one 
can use either of them to extract the properties of the percolation transition.

\bibliographystyle{apsrev4-1}
\bibliography{er-bib}

\end{document}